\documentclass[a4paper,11pt]{article}
\pdfoutput=1

\usepackage[a4paper,vmargin={30mm,30mm},hmargin={30mm,30mm}]{geometry}

\usepackage{lmodern}
\usepackage[T1]{fontenc}
\usepackage[utf8]{inputenc}

\usepackage[inline]{enumitem}

\usepackage{amsfonts}
\usepackage{amstext}
\usepackage{amsmath}
\usepackage{mathtools}
\usepackage{braket}

\usepackage{arydshln}
\usepackage[margin=\parindent,labelfont=bf]{caption}

\usepackage{pbox}

\usepackage[usenames,dvipsnames]{xcolor}

\usepackage{tikz}
\usetikzlibrary{decorations.markings}
\usetikzlibrary{arrows}

\numberwithin{equation}{section}
\numberwithin{figure}{section}

\usepackage[
colorlinks=true,
linkcolor=Red,
citecolor=Red,
urlcolor=Red,
unicode=true,
hypertexnames=false,
linktocpage=true,
bookmarksdepth=2
]{hyperref}
\hypersetup{
pdfauthor={
  Asger C. Ipsen, 
  Matthias Staudacher,
  Leonard Zippelius
}, 
pdftitle={
  title
   }
}

\usepackage{cite}
\usepackage[parsep]{collref}

\usepackage[yyyymmdd]{datetime}
\usepackage{currfile}

\usepackage[backgroundcolor=red!10!white,bordercolor=red,linecolor=red]{todonotes}

\DeclareMathOperator{\Tr}{Tr}
\begin{document} 

\baselineskip 5mm

\begin{titlepage}

  \hfill
  \pbox{5cm}{
    \texttt{HU-Mathematik-2020-01}\\
    \texttt{HU-EP-20/04}
  }
  
  \vspace{2\baselineskip}

  \begin{center}

    \textbf{\LARGE \mathversion{bold}
      The Integrable (Hyper)eclectic Spin Chain \\
    }

    \vspace{2\baselineskip}

    Changrim Ahn$^{a,b}$ and Matthias Staudacher$^{a,c}$
     
    \vspace{2\baselineskip}

\textit{
     $^a$Department of Physics, Ewha Womans University,\\
     52 Ewhayeodae-gil, Seodaemun-gu, Seoul 03760, S. Korea\\
      \vspace{0.5\baselineskip}
    }
 
    \textit{
     $^b$Korea Institute for Advanced Study (KIAS), \\
     85 Hoegiro, Dongdaemun-gu, Seoul 02455, S. Korea\\
   \vspace{0.5\baselineskip}
    }
   
     \textit{
     $^c$Institut für Mathematik und Institut für Physik, Humboldt-Universität zu Berlin,\\
      IRIS-Adlershof, Zum Großen Windkanal 6, 12489 Berlin, Germany\\
      \vspace{0.5\baselineskip}
    }
        
    \vspace{2\baselineskip}
    
     \texttt{
      ahn@ewha.ac.kr, staudacher@physik.hu-berlin.de
    }

    \vspace{2\baselineskip}

    \textbf{Abstract}
    
  \end{center}

\noindent
We refine the notion of eclectic spin chains introduced  in \cite{Ipsen:2018fmu} by including a maximal number of deformation parameters. These models are integrable, nearest-neighbor $n$-state spin chains with exceedingly simple non-hermitian Hamiltonians. They turn out to be non-diagonalizable in the multiparticle sector ($n>2$), where their ``spectrum'' consists of an intricate collection of Jordan blocks of arbitrary size and multiplicity. We show how and why the quantum inverse scattering method, sought to be universally applicable to integrable nearest-neighbor spin chains, essentially fails to reproduce the details of this spectrum. We then provide, for $n$=3,  detailed evidence by a variety of analytical and numerical techniques that the spectrum is not ``random'', but instead shows surprisingly subtle and regular patterns that moreover exhibit universality for generic deformation parameters. We also introduce a new model, the {\it hypereclectic spin chain}, where all parameters are zero except for one. Despite the extreme simplicity of its Hamiltonian, it still seems to reproduce the above ``generic'' spectra as a subset of an even more intricate overall spectrum. Our models are inspired by parts of the one-loop dilatation operator of a strongly twisted, double-scaled deformation of $\mathcal{N}=4$ Super Yang-Mills Theory.
 \end{titlepage}

{\noindent}\hrulefill

\setcounter{tocdepth}{1}
\tableofcontents{}

\vspace{1\baselineskip}
{\noindent}\hrulefill
\vspace{1\baselineskip}

%
%
%
%
\section{\mathversion{bold}Introduction and Overview}
\label{sec:intro}

The phenomenon of integrability of certain one-dimensional quantum spin chains was discovered in 1931 by Hans Bethe \cite{Bethe:1931hc}. He solved what is now known as the periodic XXX Heisenberg spin chain of length $L$, whose Hamiltonian reads 
\begin{equation}\label{eq:XXX}
\mathbf{H}=\frac{1}{2}\,\sum_{\ell=1}^L\left(1+\vec \sigma_\ell \cdot \vec \sigma_{\ell+1}\right)
\qquad {\rm with} \qquad
\vec \sigma_{\ell+1}=\vec \sigma_1\,,
\end{equation}
where $\vec \sigma_\ell$ is essentially the spin operator at site $\ell$, expressed in terms of the three Pauli matrices. $\mathbf{H}$ is a $2^L \times  2^L$ hermitian matrix acting on the tensor product space
\begin{equation}\label{eq:hilbert2}
\underbrace{
{\mathbb C}^2\otimes {\mathbb C}^2\otimes \cdots \otimes  
{\mathbb C}^2}_{L-\mbox{\scriptsize{times}}}\ .
\end{equation}
If we denote the canonical basis vectors of $\mathbb{C}^2$ by 
$| 1\, \rangle = \binom{1}{0}$ and $| 2\, \rangle = \binom{0}{1}$, the canonical basis vectors of \eqref{eq:hilbert2} are $| n_1 n_2 \ldots n_L \rangle$ with $n_j=1,2$. The Hamiltonian $\mathbf{H}$ in \eqref{eq:XXX} acts in a very transparent way on the canonical basis once one expresses it in terms of the nearest-neighbor permutation operator $\mathbb{P}^{\ell,\ell+1}$ (where $\mathbb{P}^{L,L+1}:=\mathbb{P}^{L,1}$):
\begin{equation}\label{eq:XXX2}
\mathbf{H}=\sum_{\ell=1}^L \mathbb{P}^{\ell,\ell+1}
\quad {\rm with} \quad
\mathbb{P}^{\ell,\ell+1} | \ldots n_{\ell-1}\, n_\ell\, n_{\ell+1}\,n_{\ell+2} \ldots \rangle=
| \ldots n_{\ell-1}\, n_{\ell+1}\, n_\ell\, n_{\ell+2} \ldots \rangle.
\end{equation}
Written in exactly the same form, this Hamiltonian immediately generalizes to the one of an integrable $3$-state spin chain with $n_j=1, 2, 3$ acting on 
\begin{equation}\label{eq:hilbert3}
\underbrace{
{\mathbb C}^3\otimes {\mathbb C}^3\otimes \cdots \otimes  
{\mathbb C}^3}_{L-\mbox{\scriptsize{times}}}\ .
\end{equation}
Actually, the general case of an $n$-state model, where the Hamiltonian acts on $L$ copies of $\mathbb{C}^n$, is also integrable: It takes again the same form \eqref{eq:XXX2}, except that now $n_j=1,\ldots,n$. However, in the following we will concentrate on the special case $n=3$, even though many of our findings would generalize to arbitrary $n\geq 3$. 

It is well-known that the Hamiltonian \eqref{eq:XXX2} admits, for arbitrary $n$, an integrable deformation termed {\it twisting}. It depends on up to $\frac{n(n-1)}{2}$ {\it twist parameters}. Hence, for $n=3$ one introduces three such parameters $q_1, q_2, q_3$ and replaces \eqref{eq:XXX2} with the help of a ``twisted permutation operator'' $\tilde{\mathbb{P}}^{\ell,\ell+1}_{(q_1,q_2,q_3)}$ (where still $\tilde{\mathbb{P}}^{L,L+1}_{(q_1,q_2,q_3)}:=\tilde{\mathbb{P}}^{L,1}_{(q_1,q_2,q_3)}$):
\begin{equation}\label{eq:twistedXXX}
\mathbf{\tilde{H}}_{(q_1,q_2,q_3)}=\sum_{\ell=1}^L \tilde{\mathbb{P}}^{\ell,\ell+1}_{(q_1,q_2,q_3)}\,,
\end{equation}
with
\begin{equation}\label{eq:twistedperm1}
\tilde{\mathbb{P}}^{\ell,\ell+1}_{(q_1,q_2,q_3)} | \ldots n_{\ell-1}\, n_\ell\, n_{\ell+1}\,n_{\ell+2} \ldots \rangle= 
q_{n_\ell n_{\ell+1}}
| \ldots n_{\ell-1}\, n_{\ell+1}\, n_\ell\, n_{\ell+2} \ldots \rangle,
\end{equation}
and
\begin{equation}\label{eq:twistparameters}
q_{n m}:=\delta_{n m}+\sum_{k=1}^3 |\epsilon_{n m k}|\,q_k^{-\epsilon_{n m k}}\,.
\end{equation}
Explicitly, the twisted permutation operator acts as follows on\footnote{Here we write for conciseness of notation simply $\tilde{\mathbb{P}}$ instead of  $\tilde{\mathbb{P}}^{\ell,\ell+1}_{(q_1,q_2,q_3)}$.  We also supress the labelling of the remaining $L-2$ sites, on which $\tilde{\mathbb{P}}^{\ell,\ell+1}_{(q_1,q_2,q_3)}$ acts trivially as the identity.} $|n m\rangle:= |\ldots n_\ell n_{\ell+1} \ldots \rangle$, where $n:=n_\ell$, $m:=n_{\ell+1}$ :
\begin{align}\label{eq:twistedperm2}
\tilde{\mathbb{P}}\, |11\rangle &= |11\rangle & \tilde{\mathbb{P}}\,|22\rangle &= |22\rangle & \tilde{\mathbb{P}}\,|33\rangle &= |33\rangle \\ \nonumber
\tilde{\mathbb{P}}\,|12\rangle &=\frac{1}{q_3}\, |21\rangle & \tilde{\mathbb{P}}\,|23\rangle &=\frac{1}{q_1}\, |32\rangle & \tilde{\mathbb{P}}\,|31\rangle &=\frac{1}{q_2}\, |13\rangle \\ \nonumber
\tilde{\mathbb{P}}\,|21\rangle &=q_3\, |12\rangle & \tilde{\mathbb{P}}\,|32\rangle &=q_1\, |23\rangle & \tilde{\mathbb{P}}\, |13\rangle &=q_2\, |31\rangle
\end{align}
Clearly this Hamiltonian is hermitian if and only if the parameters are complex phases, i.e.\ $q_j \in$ U$(1)$. For general complex parameters $q_j \in \mathbb{C}$ the Hamiltonian is no longer hermitian, and therefore a priori no longer diagonalizable\footnote{Numerical studies for small spin chain lengths $L$ indicate that for generic parameters $q_j \in \mathbb{C}$ the Hamiltonian \eqref{eq:twistedXXX} is nevertheless diagonalizable, albeit with complex energy eigenvalues. However, for a given $L$, one may finetune the complex parameters $q_j$ such that the Hamiltonian becomes partially non-diagonalizable (C. Ahn, M. Staudacher, unpublished).}, but still integrable. 

It is now possible to take a total of $2^3=$ {\it eight strong twisting limits} of \eqref{eq:twistedXXX} by setting
\begin{equation}\label{eq:epsilon}
q_j:=\varepsilon^{\mp1}\, \xi_j^\pm\,, \quad j=1,2,3\,,
\end{equation}
multiplying Hamiltonian and twisted permutation operators by 
$\varepsilon$, and taking  $\varepsilon$ to zero:
\begin{equation}\label{eq:stronglytwistedXXX1}
\mathbf{\hat{H}}_{(\xi_1^\pm,\xi_2^\pm,\xi_3^\pm)}=\sum_{\ell=1}^L \hat{\mathbb{P}}^{\ell,\ell+1}_{(\xi_1^\pm,\xi_2^\pm,\xi_3^\pm)}=
\lim\limits_{\varepsilon \to 0} \varepsilon\, \mathbf{\tilde{H}}_{(\varepsilon^{\mp1}\, \xi_1^\pm,\varepsilon^{\mp1}\, \xi_2^\pm, \varepsilon^{\mp1}\, \xi_3^\pm)}=
\sum_{\ell=1}^L \lim\limits_{\varepsilon \to 0}  \varepsilon\, \tilde{\mathbb{P}}^{\ell,\ell+1}_{(\varepsilon^{\mp1}\, \xi_1^\pm,\varepsilon^{\mp1}\, \xi_2^\pm,\varepsilon^{\mp1}\, \xi_3^\pm)}.
\end{equation}
For $(+,+,+)$, this leads to (with the same abbreviations as in \eqref{eq:twistedperm2})
\begin{align}\label{eq:strongtwistedperm1}
\hat{\mathbb{P}}\, |11\rangle &= 0 & \hat{\mathbb{P}}\,|22\rangle &= 0 & \hat{\mathbb{P}}\,|33\rangle &= 0 \\ \nonumber
\hat{\mathbb{P}}\,|12\rangle &=0 & \hat{\mathbb{P}}\,|23\rangle &=0 & \hat{\mathbb{P}}\,|31\rangle &=0 \\ 
\nonumber
\hat{\mathbb{P}}\,|21\rangle &=\xi_3^+\, |12\rangle & \hat{\mathbb{P}}\,|32\rangle &=\xi_1^+\, |23\rangle & \hat{\mathbb{P}}\,|13\rangle &=\xi_2^+\, |31\rangle\,,
\end{align}
while for $(-,-,-)$ we get
\begin{align}\label{eq:strongtwistedperm2}
\hat{\mathbb{P}}\, |11\rangle &= 0 & \hat{\mathbb{P}}\,|22\rangle &= 0 & \hat{\mathbb{P}}\,|33\rangle &= 0 \\ \nonumber
\hat{\mathbb{P}}\,|12\rangle &=\frac{1}{\xi_3^-}\, |21\rangle & \hat{\mathbb{P}}\,|23\rangle &=\frac{1}{\xi_1^-}\, |32\rangle & \hat{\mathbb{P}}\,|31\rangle &=\frac{1}{\xi_2^-}\, |13\rangle\, \\ 
\nonumber
\hat{\mathbb{P}}\,|21\rangle &=0 & \hat{\mathbb{P}}\,|32\rangle &=0 & \hat{\mathbb{P}}\,|13\rangle &=0\,.
\end{align}
In fact, the second case $(-,-,-)$ is equivalent to the first case $(+,+,+)$: After setting $\xi_j^-=(\xi_j^+)^{-1}$ and parity-reversing\footnote{This clearly exchanges the notion of ``right'' and ``left'': A pure convention.} the spin chain (i.e.\ replacing $|n_1 \ldots n_L\rangle$ by $|n_L \ldots n_1\rangle$) we map the second Hamiltonian to the first. These two cases are a refinement of the notion of {\it eclectic spin chain} introduced in \cite{Ipsen:2018fmu}, cf.\ section 2.3 of that paper.

The other six cases $(+,+,-)$, $(+,-,+)$, $(-,+,+)$, $(+,-,-)$,  $(-,+,-)$,  $(-,-,+)$ are different from the eclectic case. However, they are once again equivalent to each other: One checks that they are related by the six possible permutations of the three states $1,2,3$ (a mere relabelling) followed by a suitable redefinition of the twist parameters $\xi_j^\pm$. They all correspond to a refinement of the integrable spin chain model  introduced as ``broken $\mathfrak{su}(3)$ sector'' in section 5.1 of \cite{Ipsen:2018fmu}. This model is quite different from the eclectic model and will be investigated separately\footnote{C. Ahn, M. Staudacher, work in progress.}.

The eclectic three-state spin chain Hamiltonian studied in this paper will then be, after simplifying the above notation by defining $\xi_j:=\xi_j^+$, 
\begin{equation}\label{eq:stronglytwistedXXXfinal}
\mathbf{\hat{H}}_{(\xi_1,\xi_2,\xi_3)}=\sum_{\ell=1}^L \hat{\mathbb{P}}^{\ell,\ell+1}_{(\xi_1,\xi_2,\xi_3)}
\,,
\end{equation}
where
\begin{align}\label{eq:strongtwistedpermfinal}
\hat{\mathbb{P}}\, |11\rangle &= 0 & \hat{\mathbb{P}}\,|22\rangle &= 0 & \hat{\mathbb{P}}\,|33\rangle &= 0 \\ \nonumber
\hat{\mathbb{P}}\,|12\rangle &=0 & \hat{\mathbb{P}}\,|23\rangle &=0 & \hat{\mathbb{P}}\,|31\rangle &=0 \\ 
\nonumber
\hat{\mathbb{P}}\,|21\rangle &=\xi_3\, |12\rangle & \hat{\mathbb{P}}\,|32\rangle &=\xi_1\, |23\rangle & \hat{\mathbb{P}}\,|13\rangle &=\xi_2\, |31\rangle\,.
\end{align}
This leads to a novel family of, on first sight, exceedingly simple looking spin chain models with non-hermitian Hamiltonians. We will discuss their relevance and inspiration from a certain double-scaling limit of strongly twisted planar $\mathcal{N}$=4 Super-Yang-Mills theory (SYM)\cite{Gurdogan:2015csr,Sieg:2016vap,Caetano:2016ydc,Chicherin:2017cns,Gromov:2017cja,Chicherin:2017frs,Grabner:2017pgm,Kazakov:2018hrh, Gromov:2018hut} in section \ref{sec:origin}. We will also verify the integrability of these models for arbitrary complex parameters $\xi_j$ in \ref{sec:integrability}. However, except for trivial sectors, where the states are made up from only two of the three excitations, these Hamiltonians are {\it nilpotent} on any state containing all three excitations \cite{Ipsen:2018fmu}. They are therefore totally non-diagonalizable. The most one can do is to bring them into {\it Jordan normal form} (JNF). However, as we will outline in section \ref{sec:failure} and then demonstrate in more detail in chapter \ref{sec:QISM}, the quantum inverse scattering method mostly fails\footnote{It does account for the fact that the {\it generalized energy eigenvalues} of the Jordan blocks are all zero.}, at least in its traditional form, to be helpful for this enterprise. This is vexing, in particular since we shall find, through some case-by-case numerical studies in chapter \ref{sec:results} up to moderately large lengths $L$, that the sizes and multiplicities of the appearing Jordan block show intriguingly regular patterns. We shall also find evidence for a certain {\it universality} of the ``spectrum'' of Jordan blocks: Its dependence on complex parameters $\xi_j$ is relatively weak, in a sense to be explained, as long as these are suitably ``generic''.

A non-generic situation arises e.g.\ when some of the parameters $\xi_j$ are zero (which is allowed!). The most extreme case is that two of them are zero, say $\xi_1=\xi_2=0$. 
We may then set without loss of generality $\xi_3=1$ to obtain (again with $\mathfrak{P}^{L,L+1}=\mathfrak{P}^{L,1}$)
\begin{equation}\label{eq:hypereclecticH}
\mathfrak{H}=\sum_{\ell=1}^L \mathfrak{P}^{\ell,\ell+1}
\,,
\end{equation}
where
\begin{align}\label{eq:hypereclecticP}
\mathfrak{P}\, |11\rangle &= 0 & \mathfrak{P}\,|22\rangle &= 0 & \mathfrak{P}\,|33\rangle &= 0 \\ \nonumber
\mathfrak{P}\,|12\rangle &= 0 & \mathfrak{P}\,|23\rangle &=0 & \mathfrak{P}\,|31\rangle &=0 \\ 
\nonumber
\mathfrak{P}\,|21\rangle &=|12\rangle & \mathfrak{P}\,|32\rangle &=0 & \mathfrak{P}\,|13\rangle &=0\,.
\end{align}
This novel system must surely be the simplest one among all integrable three-state spin chains. We call it the {\it hypereclectic model}. While looking trivial on first sight, we will demonstrate in chapter \ref{sec:results} that its intricate ``spectrum'' of Jordan blocks is actually even richer than the one of the more general and more complicated looking model \eqref{eq:stronglytwistedXXXfinal}, \eqref{eq:strongtwistedpermfinal} for generic parameters $\xi_j$.
\section{\mathversion{bold}Integrability, Origin, Non-Diagonalizability of the Models}
\label{sec:model}
\subsection{\mathversion{bold}Quantum Integrability of the Eclectic Spin Chain Models}
\label{sec:integrability}
Let us begin by establishing the quantum integrability of the spin chain Hamiltonians defined in chapter~\ref{sec:intro}. We will assume some basic familiarity with the quantum inverse scattering method, and in particular with the Algebraic Bethe Ansatz (ABA) technique, see e.g.\ \cite{Faddeev:1996iy, Nepomechie:1998jf} for excellent introductions. Since we are discussing three-state models, an additional complication is the necessity to consider a {\it nested} ABA. For a recent pedagogical introduction see e.g.\ \cite{Levkovich-Maslyuk:2016kfv}. The starting point is an R-matrix of size, in the three-state case, $9 \times 9$ that acts on the tensor product $\mathbb{C}^3 \otimes \mathbb{C}^3$. In the twisted case it reads
\begin{equation}\label{eq:Rmat}
\mathbf{\tilde{R}}_{(q_1,q_2,q_3)}^{12}(u)=\left(
\begin{array}{ccc|ccc|ccc}
u+1& & & & & & & & \\
& \frac{u}{q_3}& &1 & & & & & \\
& & q_2\, u& & & & 1& & \\
\hline
& 1& &q_3\, u & & & & & \\
& & & &u+1 & & & & \\
& & & & & \frac{u}{q_1} & & 1& \\
\hline
& &1 & & & &\frac{u}{q_2}  & & \\
& & & & & 1& &q_1\, u & \\
& & & & & & & & u+1
 \end{array}\right),
 \end{equation}
where $u$ is termed {\it spectral parameter}, and the upper indices $1,2$ label the two copies of the space $\mathbb{C}^3$ in the tensor product. One now uses this R-matrix as a so-called {\it Lax-operator}\footnote{For the expert reader we remark that the usual shift of the spectral parameter, that, by convention, marks the difference between the R-matrix and the Lax-operator, is not suitable for our purposes.} and builds up a {\it quantum monodromy matrix}
\begin{equation}\label{eq:monodromy}
\mathbf{\tilde{M}}_{(q_1,q_2,q_3)}^{a,L}(u)=
\mathbf{\tilde{R}}_{(q_1,q_2,q_3)}^{a,L}(u)\cdot \mathbf{\tilde{R}}_{(q_1,q_2,q_3)}^{a,L-1}(u)\cdot \ldots
\mathbf{\tilde{R}}_{(q_1,q_2,q_3)}^{a,2}(u)\cdot \mathbf{\tilde{R}}_{(q_1,q_2,q_3)}^{a,1}(u).
\end{equation}
Here $\cdot$ denotes $3 \times 3$ matrix multiplication in an {\it auxiliary space}, which is another copy of $\mathbb{C}^3$. The entries of this $3 \times 3$ matrix act on \eqref{eq:hilbert3}. Therefore, 
$\mathbf{\tilde{M}}_{(q_1,q_2,q_3)}^{a,L}(u)$ acts on the tensor product of \eqref{eq:hilbert3} and the auxiliary space $a$. It is customary to drop the index $L$. One then takes the trace over the space $a$, and obtains the {\it transfer matrix} acting on \eqref{eq:hilbert3}:
\begin{equation}\label{eq:transfer}
\mathbf{\tilde{T}}_{(q_1,q_2,q_3)}(u)={\rm Tr}_a\, \mathbf{\tilde{M}}_{(q_1,q_2,q_3)}^a(u).
 \end{equation}
 It is well-known (e.g.\ \cite{Faddeev:1996iy, Nepomechie:1998jf}) that it yields for $u=0$ the shift operator of the spin chain, i.e.
 \begin{equation}\label{eq:shift}
\mathbf{U}=\mathbf{\tilde{T}}_{(q_1,q_2,q_3)}(0)\,,
\quad {\rm where} \quad
\mathbf{U}\,|n_1 n_2 \ldots n_{L-1} n_L \rangle=|n_L n_1 \ldots n_{L-2} n_{L-1} \rangle.
 \end{equation}
 It is also known that a nearest-neighbor Hamiltonian may be extracted as the logarithmic derivative of the transfer matrix at $u=0$, and for \eqref{eq:Rmat} one finds precisely \eqref{eq:twistedXXX}:
\begin{equation}\label{eq:hamil}
\mathbf{\tilde{H}}_{(q_1,q_2,q_3)}=\mathbf{U}^{-1}\,\frac{d}{du}\,\mathbf{\tilde{T}}_{(q_1,q_2,q_3)}(u)\bigg|_{u=0}\,.
 \end{equation}
 The R-matrix \eqref{eq:Rmat} of the twisted model satisfies the {\it Yang-Baxter equation}
\begin{equation}\label{eq:ybe}
\mathbf{\tilde{R}}_{(q_1,q_2,q_3)}^{12}(u-u')\mathbf{\tilde{R}}_{(q_1,q_2,q_3)}^{13}(u)\mathbf{\tilde{R}}_{(q_1,q_2,q_3)}^{23}(u')
=
\mathbf{\tilde{R}}_{(q_1,q_2,q_3)}^{23}(u')\mathbf{\tilde{R}}_{(q_1,q_2,q_3)}^{13}(u)\mathbf{\tilde{R}}_{(q_1,q_2,q_3)}^{12}(u-u')
\end{equation}
for arbitrary complex values of $q_1,q_2,q_3$. This implies \cite{Faddeev:1996iy} the following relation intertwining the monodromy matrices $\mathbf{\tilde{M}}_{(q_1,q_2,q_3)}^a$ with the R-matrix $\mathbf{\tilde{R}}_{(q_1,q_2,q_3)}^{12}$:
\begin{equation}\label{eq:RTT}
\mathbf{\tilde{R}}_{(q_1,q_2,q_3)}^{12}(u-u')\mathbf{\tilde{M}}_{(q_1,q_2,q_3)}^1(u)\mathbf{\tilde{M}}_{(q_1,q_2,q_3)}^2(u')
=
\mathbf{\tilde{M}}_{(q_1,q_2,q_3)}^2(u')\mathbf{\tilde{M}}_{(q_1,q_2,q_3)}^1(u)\mathbf{\tilde{R}}_{(q_1,q_2,q_3)}^{12}(u-u')
\end{equation}
Taking the traces over the spaces labelled by $1$ and $2$, one derives the following commutation relations for arbitrary complex values $u,u'$:
\begin{equation}\label{eq:commutation}
[\mathbf{\tilde{T}}_{(q_1,q_2,q_3)}(u),\mathbf{\tilde{T}}_{(q_1,q_2,q_3)}(u')]=0
\quad {\rm and}\,\,  {\rm thence} \quad
[\mathbf{\tilde{H}}_{(q_1,q_2,q_3)},\mathbf{\tilde{T}}_{(q_1,q_2,q_3)}(u')]=0\,.
\end{equation}
Since the transfer matrix may be interpreted as a generating function of a sufficiently large number of independent charges in involution, and since the Hamiltonian is commuting with these, one calls, by definition, the model {\it quantum integrable}.

Let us now consider the strong twisting limits of the above R-matrices. The upshot is, that quantum integrability as defined above survives the limiting procedure. Here we will only consider the limit to the eclectic spin chain model, as discussed in chapter \ref{sec:intro}: We multiply the R-matrix \eqref{eq:Rmat} by $\varepsilon$, put $q_i:=\varepsilon^{-1}\, \xi_j$, and replace $u$ by $\varepsilon\,u$. We then take  
$\varepsilon$ to zero. This results in
\begin{equation}\label{eq:scaledRmat}
\mathbf{\hat{R}}_{(\xi_1,\xi_2,\xi_3)}(u)=\left(
\begin{array}{ccc|ccc|ccc}
1& & & & & & & & \\
& & &1 & & & & & \\
& & \xi_2\, u& & & & 1& & \\
\hline
& 1& &\xi_3\, u & & & & & \\
& & & &1 & & & & \\
& & & & & & & 1& \\
\hline
& &1 & & & & & & \\
& & & & & 1& &\xi_1\, u & \\
& & & & & & & & 1
 \end{array}
 \right).
\end{equation}
Taking the logarithmic derivative at $u=0$ as in \eqref{eq:hamil} results precisely in the Hamiltonian 
$\mathbf{\hat{H}}_{(\xi_1,\xi_2,\xi_3)}$ in \eqref{eq:stronglytwistedXXXfinal},\eqref{eq:strongtwistedpermfinal}. It is still integrable, as one easily checks that the Yang-Baxter equation \eqref{eq:ybe} also holds with $\mathbf{\tilde{R}}_{(q_1,q_2,q_3)}$ replaced by the eclectic R-matrix $\mathbf{\hat{R}}_{(\xi_1,\xi_2,\xi_3)}$. For completeness we also state the R-matrix for the hypereclectic model defined in chapter \ref{sec:intro}:
\begin{equation}\label{eq:hyperRmat}
\mathbf{\mathfrak{R}}(u)=\left(
\begin{array}{ccc|ccc|ccc}
1& & & & & & & & \\
& & &1 & & & & & \\
& & & & & & 1& & \\
\hline
& 1& & u & & & & & \\
& & & &1 & & & & \\
& & & & & & & 1& \\
\hline
& &1 & & & & & & \\
& & & & & 1& & & \\
& & & & & & & & 1
 \end{array}
 \right).
\end{equation}
It also satisfies the Yang-Baxter equation and generates the integrable hypereclectic Hamiltonian 
$\mathfrak{H}$ in \eqref{eq:hypereclecticH},\eqref{eq:hypereclecticP}.
\subsection{Interlude: Origin and Relevance of the Models}
\label{sec:origin}
    Let us briefly pause to discuss the origin of the models introduced in chapter \ref{sec:intro} and section \ref{sec:integrability}. Applying twisted boundary conditions to quantum spin chains has a long history in condensed matter theory. They are of physical interest, preserve integrability as we just recalled, and yield a better understanding of the inner workings of the quantum inverse scattering method, to be discussed below. However, twisted integrable spin chains are also of relevance in certain conformal quantum field theories in four (as well as in three and two) dimensions, see \cite{Beisert:2010jr}, and in particular \cite{Zoubos:2010kh}, where their relation to the so-called three-parameter $\gamma_j$-deformed versions of ${\cal N}=4$ SYM is explained. This was the starting point for a series of papers \cite{Gurdogan:2015csr,Sieg:2016vap,Caetano:2016ydc,Chicherin:2017cns,Gromov:2017cja,Chicherin:2017frs,Grabner:2017pgm,Kazakov:2018hrh, Gromov:2018hut} that introduced and started to analyze a novel class of integrable conformal four-dimensional quantum field theories. Let us recall the double-scaling limit used in \cite{Gurdogan:2015csr,Sieg:2016vap,Caetano:2016ydc,Chicherin:2017cns,Gromov:2017cja,Chicherin:2017frs,Grabner:2017pgm,Kazakov:2018hrh, Gromov:2018hut}:
Defining the square of the planar gauge theory coupling constant as $g^2=\frac{\lambda}{16 \pi^2}$, where $\lambda$ is the `t~Hooft coupling, it was suggested to take $g\rightarrow 0$, while some or all of the twisting parameters either turn to $q_j=e^{-i \gamma_j /2}\rightarrow \infty$ or else to $q_j=e^{-i \gamma_j /2}\rightarrow 0$, such that, respectively, either the products $g\, q_j$ or quotients 
$g\, q_j^{-1}$ are held fixed. As in \cite{Ipsen:2018fmu}, we shall call these double-scaled, twisted models simply ``strongly twisted models''.     
A neat way to systematically treat this total of $2^3=8$ different strong twisting limits of  $\gamma_j$-deformed versions of ${\cal N}=4$ SYM is to use the parameter $\varepsilon$ introduced in \eqref{eq:epsilon}: Write $q_j:=\varepsilon^{\mp1}\, \xi_j^\pm$, replace $g \rightarrow \varepsilon\, g$, and take $\varepsilon$ to zero. This results in, respectively, the limits $g\, q_j\rightarrow g\, \xi_j^+$ and $g\, q_j^{-1}\rightarrow g\, (\xi_j^-)^{-1}$. In all eight cases, the gauge fields decouples from the interacting part of the Lagrangian. As in \cite{Gurdogan:2015csr,Sieg:2016vap,Caetano:2016ydc,Chicherin:2017cns,Gromov:2017cja,Chicherin:2017frs,Grabner:2017pgm,Kazakov:2018hrh, Gromov:2018hut}, we will simply ignore them altogether\footnote{Taking the double-scaling limits on the level of the twisted Lagrangian of $\gamma_j$-deformed ${\cal N}=4$ SYM, the gauge fields are still present in the kinetic terms, of course. However, if we choose to only consider local composite operators not containing gauge fields, they indeed completely decouple: Notice that the gauge fields also decouple from all covariant derivatives, turning them into ordinary ones.}. The limit where all three couplings are scaled as $g\, q_j\rightarrow g\, \xi_j^+$ (i.e.\ all $q_{1,2,3} \rightarrow \infty$) reduces the interacting part of the $\gamma_j$-deformed Lagrangian of ${\cal N}=4$ SYM to 
\begin{eqnarray}
\mathcal{L}_\text{int}(g,\xi_1^+,\xi_2^+,\xi_3^+)&=&-g^2\, N\Tr\left( (\xi_3^+)^2\, \phi_1^\dagger \phi_2^\dagger \phi^1 \phi^2  +  (\xi_2^+)^2\, \phi_3^\dagger \phi_1^\dagger \phi^3 \phi^1  +  (\xi_1^+)^2\, \phi_2^\dagger \phi_3^\dagger \phi^2 \phi^3\right) 
\nonumber \\  
&& -g\, N \Tr\left( i\, \sqrt{\xi_2^+ \xi_3^+} (    \psi^3\phi^1\psi^2+ \bar\psi_3 \phi_1^\dagger \bar\psi_2) + \text{cyclic} \right).
\label{eq:Lagrangian1}
\end{eqnarray}
where by ``cyclic'' we mean cyclic permutations of the three indices. We do not show the standard kinetic terms of the complex bosonic and fermionic fields. Note that the fourth fermion $\psi_4$ decouples from the interactive part. Conversely, in the limit $g\, q_j^{-1}\rightarrow g\, (\xi_j^-)^{-1}$ 
(i.e.\ all $q_{1,2,3} \rightarrow 0$) we obtain
\begin{eqnarray}
\mathcal{L}_\text{int}(\xi_1^-,\xi_2^-,\xi_3^-)&=&
-g^2\, N\Tr\left( (\xi_3^-)^{-2}\, \phi_2^\dagger \phi_1^\dagger \phi^2 \phi^1  
+  (\xi_2^-)^{-2}\, \phi_1^\dagger \phi_3^\dagger \phi^1 \phi^3  
+  (\xi_1^-)^{-2}\, \phi_3^\dagger \phi_2^\dagger \phi^3 \phi^2\right)  \nonumber \\ 
&& -g\, N \Tr\left( \frac{i}{\sqrt{\xi_2^- \xi_3^-}} (    \psi^2\phi^1\psi^3
+ \bar\psi_2 \phi_1^\dagger \bar\psi_3) + \text{cyclic} \right).
\label{eq:Lagrangian2}
\end{eqnarray}
The other six limits are more subtle. If e.g.\ we take the limits 
$g, q_3 \rightarrow 0$, $q_{1,2} \rightarrow \infty$, i.e.\
\begin{equation}
g\, q_1\rightarrow g\, \xi_1^+\,,
\qquad
g\, q_2\rightarrow g\, \xi_2^+\,,
\qquad
g\, q_3^{-1}\rightarrow g\, (\xi_3^-)^{-1}\,, 
\end{equation}
we end up (setting for simplicity the redundant overall coupling $g=1$) with
\begin{equation}\label{eq:Lagrangian3}
\mathcal{L}_\text{int}(\xi_1^+,\xi_2^+,\xi_3^-)=
- N\Tr\Big( (\xi_3^-)^{-2}\, \phi_2^\dagger \phi_1^\dagger \phi^2 \phi^1  
+  (\xi_2^+)^2\, \phi_3^\dagger \phi_1^\dagger \phi^3 \phi^1  
+  (\xi_1^+)^2\, \phi_2^\dagger \phi_3^\dagger \phi^2 \phi^3  
\end{equation}
\begin{equation}
+\sqrt{\frac{\xi_2^+}{\xi_3^-}} 
\left( \bar \psi_1 \phi^1 \bar \psi_4 - \psi^1 \phi_1^\dagger \psi^4 \right)
- \sqrt{\frac{\xi_1^+}{\xi_3^-}}
\left( \bar \psi_4 \phi^2 \bar \psi_2 - \psi^4 \phi_2^\dagger \psi^2 \right)
-i \sqrt{\xi_1^+ \xi_2^+} \left(\bar \psi_2 \phi_3^\dagger \bar \psi_1 + \psi^2 \phi^3 \psi^1 \right)
\Big). \nonumber
\end{equation}
If we put $\xi_1^+=0$, $\xi_2^+=\xi_2$ and $\xi_3^-=\xi_3^{-1}$ we recover as a special case the model mentioned in equations (8),(9) of \cite{Caetano:2016ydc}.

Now, as argued in \cite{Gurdogan:2015csr,Sieg:2016vap,Caetano:2016ydc} and explained in more detail in \cite{Ipsen:2018fmu}, the one-loop dilatation operator of the above models is closely related to the integrable Hamiltonians discussed in chapter \ref{sec:intro} and section \ref{sec:integrability}. The precise statement is as follows: If we analyze the strongly twisted quantum field theories (QFTs) with the above Lagrangians, and choose to only study local composite operators\footnote{
That is, we do not consider operators containing any derivatives, fermions, anti-chiral fields $\phi_1^\dagger, \phi_2^\dagger, \phi_3^\dagger$, nor the (in any case decoupled) gauge fields.}
containing the three partonic chiral fields $\phi_1, \phi_2, \phi_3$, the QFTs 
dilatation operator $\mathfrak{D}$  is given by 
\begin{equation}\label{eq:dila}
\mathfrak{D}=\mathfrak{D}_0+g^2\,\mathbf{\hat{H}}_{(\xi_1^\pm,\xi_2^\pm,\xi_3^\pm)}+\mathcal{O}(g^4),
\end{equation}
where $\mathfrak{D}_0$ is the classical dilatation operator\footnote{
Here $\mathfrak{D}_0$ only counts the total number of $\phi$'s of the composite operator = length $L$ of the spin chain.},
and the one-loop dilatation operator is $g^2$ times the spin chain Hamiltonian 
$\mathbf{\hat{H}}_{(\xi_1^\pm,\xi_2^\pm,\xi_3^\pm)}$ in \eqref{eq:stronglytwistedXXX1}. Once again, in this paper we focus on the eclectic $(+,+,+)$ case with $\xi_j:=\xi_j^+$, where the Hamiltonian is $\mathbf{\hat{H}}_{(\xi_1,\xi_2,\xi_3)}$ in \eqref{eq:stronglytwistedXXXfinal}, \eqref{eq:strongtwistedpermfinal}, with extra attention given to the special hypereclectic case, where $\xi_1=\xi_2=0$, $\xi_3=1$ with Hamiltonian $\mathfrak{H}$ in 
\eqref{eq:hypereclecticH}, \eqref{eq:hypereclecticP}. The gauge theory interpretation of this model is as follows. The one-loop dilatation operator still acts on composite operators containing the three chiral fields $\phi_1, \phi_2, \phi_3$, which are to be interpreted as the one-site states $1$, $2$, $3$ in the spin chain interpretation, respectively. 
However, now the dilatation operator merely chirally exchanges $\phi_1, \phi_2$, while $\phi_3$ becomes a non-moving ``spectator field'':
\begin{equation}\label{eq:dilahypereclectic}
\mathfrak{D}=\mathfrak{D}_0+g^2\,\mathfrak{H}+\mathcal{O}(g^4).
\end{equation}
As we shall see below, these spectator fields act as impenetrable domain walls within the spin chain, leading to very intricate effects.

For the remainder of this paper we will ignore the gauge theory origin and interpretation of the integrable eclectic and hypereclectic spin chains, and simply study their intricate ``spectrum'' on its own right. We shall begin by explaining in the next section \ref{sec:failure} why and how the spectral problem for diagonalizable spin chains is to be replaced, in the case of the non-diagonalizable eclectic and hypereclectic chains, with the problem of finding the appropriate JNF of the transfer matrix and the Hamiltonian. Correspondingly, we shall furthermore demonstrate how and why the direct application of the ABA to the integrable spin chains built from the R-matrix \eqref{eq:scaledRmat}, \eqref{eq:hyperRmat} is bound to fail.
\subsection{Non-Diagonalizability, Jordan blocks, and Failure of the ABA}
\label{sec:failure}

For generic, finite, complex twist parameters $q_i$ the transfer matrix 
$\mathbf{\tilde{T}}_{(q_1,q_2,q_3)}(u)$ in \eqref{eq:transfer} of the three-state spin chain is diagonalizable with $3^L$ eigenvalues $\tilde{T}^j_{(q_1,q_2,q_3)(u)}$, $j=1,\ldots ,3^L$:
\begin{equation}\label{eq:transferev}
\mathbf{\tilde{T}}_{(q_1,q_2,q_3)}(u)\, |\psi_j\rangle=\tilde{T}^j_{(q_1,q_2,q_3)}(u)\, |\psi_j\rangle.
 \end{equation}
Because of \eqref{eq:commutation} the linearly independent eigenstates $| \psi_j \rangle$ are also eigenstates\footnote{
In gauge theory applications we are only interested in the subset of cyclic states with $k_j=0$.
} 
of the shift operator $\mathbf{U}=\mathbf{\tilde{T}}_{(q_1,q_2,q_3)}(0)$ in \eqref{eq:shift2}.
Defining
\begin{equation}\label{eq:rootfunity}
\omega_L:=e^{\frac{2 \pi i}{L}}\,.
\end{equation}
one has, from the condition $\mathbf{U}^L=\mathbb{I}$,
\begin{equation}\label{eq:shiftev}
\mathbf{U}\, |\psi_j\rangle=\omega_L^{k_j}\, |\psi_j\rangle
\qquad {\rm with} \qquad
k_j\in \{0, \ldots, L-1\}\,.
 \end{equation}
In light of \eqref{eq:hamil}, the $|\psi_j\rangle$ are of course also eigenstates of the Hamiltonian $\mathbf{\tilde{H}}_{(q_1,q_2,q_3)}$:
\begin{equation}\label{eq:energyev}
\mathbf{\tilde{H}}_{(q_1,q_2,q_3)}\, |\psi_j\rangle=\tilde E^j_{(q_1,q_2,q_3)}\, |\psi_j\rangle.
 \end{equation}
Integrability allows to find this spectrum, for example by means of the nested ABA (see  e.g.\ \cite{Levkovich-Maslyuk:2016kfv}). The ansatz for an eigenstate with $L-M$ states $1$, $M-K$ states $2$, and $K$ states $3$ is
\begin{equation}\label{eq:ABA1}
\vert\psi_j\rangle=\sum_{\{a_i\}=1,2}\mathbf{F}^{a_1\cdots a_M}\mathcal{B}_{a_1}(u_1)\cdots
\mathcal{B}_{a_M}(u_M)\vert \underbrace{11\cdots 1}_L\,\rangle,
\end{equation}
where the creation operators $\mathcal{B}_{a}$ ($a=1,2$) act on the so-called reference state 
$\vert 11\cdots 1\rangle$, which is exclusively made of fields of type $1$ at all $L$ sites. Each $\mathcal{B}_{1}$ or $\mathcal{B}_{2}$ creates a linear combination of new states, where one of the $1$'s is replaced by a $2$ or $3$, respectively. The $\mathcal{B}_{1}(u)$, $\mathcal{B}_{2}(u)$ are obtained as two of the components of the monodromy matrix \eqref{eq:monodromy}, written as a $3 \times 3$ operator-valued matrix in auxiliary space $a$ as
\begin{equation}\label{eq:monodromy2}
\mathbf{\tilde{M}}_{(q_1,q_2,q_3)}^a(u)=
\left(
\begin{array}{c|cc}
\mathcal{A}(u)& \mathcal{B}_1(u)& \mathcal{B}_2(u)\\
\hline
\mathcal{C}_1(u)& \mathcal{D}_{11}(u)&\mathcal{D}_{12}(u)\\
\mathcal{C}_2(u)& \mathcal{D}_{21}(u)&\mathcal{D}_{22}(u)
\end{array}
\right).
\end{equation}
In the nested Bethe ansatz, $\mathbf{F}^{a_1\cdots a_M}$ is an eigenstate of a secondary, inhomogeneous rank-one spin-chain of length $M$.
It may be constructed by acting $K$ times with the rank-one creation operators
$\mathbf{b}(v_i)$ on a pseudo-vacuum with inhomogeneities $u_1, \ldots, u_M$:
\begin{equation}\label{eq:ABA2}
 \vert\mathbf{F}^{a_1\cdots a_M}\mathcal{B}_{a_1}(u_1)\cdots
 \mathcal{B}_{a_M}(u_M)\rangle=\mathbf{b}(v_1)\cdots\mathbf{b}(v_K)\vert
 \mathcal{B}_{1}(u_1)\cdots
 \mathcal{B}_{1}(u_M)\rangle.
\end{equation}
The ansatz \eqref{eq:ABA1} works, i.e.\ $\vert\psi_j\rangle$ indeed becomes an eigenstate of the transfer matrix, if and only if both the {\it level-one Bethe roots} $u_1, \ldots, u_M$ as well as the {\it level-two Bethe roots} $v_1, \ldots, u_K$ are meticulously chosen as solutions of a nested Bethe ansatz, see section \ref{sec:ABA} below. 
The transfer matrix eigenvalues $\tilde{T}^j_{(q_1,q_2,q_3)}(u)$ are then determined explicitly through the level-one Bethe roots. This procedure is known to yield, as long as the $q_i$ are generic, the {\it complete} spectrum of $3^L$ eigenstates, once all solutions are of the BAE are found.

In contrast, the eclectic model is very different. Let us already summarize the gist of our findings (for more details, see the following two sections). For the Hamiltonian $\mathbf{\hat{H}}_{(\xi_1,\xi_2,\xi_3)}$ of \eqref{eq:stronglytwistedXXXfinal}, \eqref{eq:strongtwistedpermfinal} one has to replace \eqref{eq:energyev} by
\begin{equation}\label{eq:energygenev}
\left( \mathbf{\hat{H}}_{(\xi_1,\xi_2,\xi_3)} - \hat E^j_{(\xi_1,\xi_2,\xi_3)}\right)^{m_j}\, |\psi_j^{m_j}\rangle=0
\qquad {\rm with} \qquad
m_j = 1, \ldots, l_j\,.
\end{equation}
Here the  $|\psi_j^{m_j}\rangle$ are so-called {\it generalized eigenstates}, associated to {\it generalized eigenvalues} $E^j_{(\xi_1,\xi_2,\xi_3)}$. Let us define a 
%
%
{\it Jordan block} $J_l(\lambda)$ (or {\it Jordan matrix}) as a matrix of size $l \times l$ for $\lambda \in \mathbb{C}$ by
\begin{equation}\label{eq:JordanBlock}
J_l(\lambda):=\left(
\begin{array}{ccccc}
 \lambda & 1 &    &            &  0 \\
    & \lambda & 1 &            &      \\
    &    & \lambda & \ddots &     \\
    &    &    & \ddots &  1 \\
 0 &    &    &            &  \lambda 
 \end{array}
\right).
\end{equation}
The statement is then that, for a non-diagonalizable $3^L \times 3^L$ matrix such as the Hamiltonian $\mathbf{\hat{H}}_{(\xi_1,\xi_2,\xi_3)}$, the best one can do is to bring it, by a suitable $3^L \times 3^L$ similarity transform $S$, into JNF
\begin{equation}\label{eq:energyJNF}
S \cdot \mathbf{\hat{H}}_{(\xi_1,\xi_2,\xi_3)} \cdot S^{-1}=\left(
\begin{array}{ccc}
 J_{l_1}\left(\hat E^1_{(\xi_1,\xi_2,\xi_3)}\right) &  &      0 \\
    & \ddots &      \\
 0   &    &  J_{l_b}\left(\hat E^b_{(\xi_1,\xi_2,\xi_3)}\right)   
  \end{array}
\right),
\end{equation}
where $b$ is the total number of Jordan blocks, and their sizes add up to
\begin{equation}\label{eq:JordanCompleteness}
l_1 + \ldots + l_b=3^L\,. 
\end{equation}
In JNF, \eqref{eq:energygenev} is refined, with $|\varphi_j^{m_j}\rangle=S\,|\psi_j^{m_j}\rangle$, into
\begin{equation}\label{eq:energyJNF1}
S \cdot \left( \mathbf{\hat{H}}_{(\xi_1,\xi_2,\xi_3)} - \hat E^j_{(\xi_1,\xi_2,\xi_3)}\right) \cdot S^{-1}\, 
|\varphi_j^{m_j}\rangle=|\varphi_j^{m_j-1}\rangle
\quad {\rm for} \quad
m_j = 2, \ldots, l_j
\end{equation}
while for $m_j = 1$ one has 
\begin{equation}\label{eq:energyJNF2}
S \cdot \left( \mathbf{\hat{H}}_{(\xi_1,\xi_2,\xi_3)} - \hat E^j_{(\xi_1,\xi_2,\xi_3)}\right)\cdot S^{-1}\, |\varphi_j^1\rangle=0\,.
\end{equation}
The last equation shows that for each Jordan block, labeled by $j$, there is exactly one true (i.e.\ non-generalized) eigenstate $|\varphi_j^1\rangle$ with eigenvalue $\hat E^j_{(\xi_1,\xi_2,\xi_3)}$.

The two-state sectors, where the matrix acts on spin chains that only contain two out of the three fields, are diagonalizable, i.e.\ the corresponding ``Jordan blocks'' are all of size $l_j=1$, and their exact spectrum is easily found \cite{Ipsen:2018fmu}. However, in the three-state sectors there are many Jordan blocks of size $l_j>1$, even though some $l_j=1$ blocks remain. In fact, as already argued in \cite{Ipsen:2018fmu}, the generalized eigenvalues are all zero: 
\begin{equation}\label{eq:energygeneval}
E^j_{(\xi_1,\xi_2,\xi_3)}=0\,.
\end{equation}
The Hamiltonian becomes nilpotent of degree $l_j$, see \eqref{eq:energygenev}. Note however that the generalized eigenstates are still ordinary eigenstates of the shift operator $\mathbf{U}=\mathbf{\tilde{T}}_{(q_1,q_2,q_3)}(0)$ (compare to \eqref{eq:shiftev}):
\begin{equation}\label{eq:shiftgenev}
\mathbf{U}\, |\varphi_j^{m_j}\rangle=\omega_L^{k_j}\, |\varphi_j^{m_j}\rangle
\qquad {\rm with} \qquad
k_j\in \{0, \ldots, L-1\}
 \end{equation}
for all $m_j = 1, \ldots, l_j$. 
Nevertheless, the transfer matrix $\mathbf{\tilde{T}}_{(q_1,q_2,q_3)}(u)$ is {\it only} diagonalizable at $u=0$. As we shall see below, for $u \neq 0$ the situation is analogous to the one for the Hamiltonian: \eqref{eq:transferev} only holds in the two-particle sectors, while in general one has the same situation as in \eqref{eq:energygenev}:
\begin{equation}\label{eq:transfergenev}
\left( \mathbf{\hat{T}}_{(\xi_1,\xi_2,\xi_3)}(u) - \hat T^j_{(\xi_1,\xi_2,\xi_3)}(u)\right)^{m_j}\, |\psi_j^{m_j}\rangle=0
\qquad {\rm with} \qquad
m_j = 1, \ldots, l_j
 \end{equation}
for {\it generalized eigenstates} $|\psi_j^m\rangle$. Surprisingly, in the three-particle\footnote{
In the diagonalizable ($l_j=1$) 2-particle sectors the $\hat T^j_{(\xi_1,\xi_2,\xi_3)}(u)$ are instead known polynomials in $u$.
}
sectors of the eclectic model the {\it generalized eigenvalues} turn out to be $u$-independent (incidentally explaining via \eqref{eq:hamil} the result \eqref{eq:energygeneval}), and are in fact equal to the eigenvalues of the shift operator:
\begin{equation}\label{eq:transfergeneval}
\hat T^j_{(\xi_1,\xi_2,\xi_3)}(u)= \omega_L^{k_j}
\qquad {\rm with} \qquad
k_j\in \{0, \ldots, L-1\}\,.
\end{equation}
Just like \eqref{eq:energyJNF}, the transfer matrix may be brought by a suitable $3^L \times 3^L$ similarity transform $S'(u)$ into the JNF\footnote{
The Jordan block structure of the transfer matrix \eqref{eq:transfergeneval} a priori does not have to coincide with the one of the Hamiltonian, \eqref{eq:energyJNF}. However, we suspect that the two Jordan decompositions coincide. We have found no counterexamples so far, but also could not yet find a proof of this conjecture.
}
\begin{equation}\label{eq:transferJNF}
S'(u) \cdot \mathbf{\hat{T}}_{(\xi_1,\xi_2,\xi_3)}(u) \cdot S'(u)^{-1}=\left(
\begin{array}{ccc}
 J_{l_1}\left(\hat T^1_{(\xi_1,\xi_2,\xi_3)}(u)\right) &  &      0 \\
    & \ddots &      \\
 0   &    &  J_{l_b}\left(\hat T^b_{(\xi_1,\xi_2,\xi_3)}(u)\right)   
  \end{array}
\right),
\end{equation}
with $l_1 + \ldots + l_b=3^L$. Note that $S'(u)$ is $u$-dependent and as such different from $S$ in
\eqref{eq:energyJNF}. This is in stark contrast to ordinary integrable spin chains, where the similarity transform that diagonalizes the transfer matrix is $u$-independent and simultaneously diagonalizes all commuting charges including the Hamiltonian, cf.\ \eqref{eq:commutation}.

Given this structure, one would hope that a generalization of the ABA exists that would allow the construction of the generalized eigenstates $|\psi_j^m\rangle$, in order to extract the number and sizes of Jordan blocks of the eclectic model. Unfortunately, we have not yet been able to find such a generalization. The first and foremost problem seems to be that the {\it ansatz} \eqref{eq:ABA1} simply fails from the very start: For all values of the $\mathbf{F}^{a_1\cdots a_M}$ and $u_1, \ldots, u_M$ the ansatz does not span the full state space needed to build the wanted generalized eigenstates. Disappointingly, for the majority of Jordan blocks $j$, it  does not even account for the $m_j=1$ states of the block, which {\it are} ordinary eigenstates\footnote{
Every Jordan block possesses exactly one non-generalized, ordinary eigenstate, the state with $m_j=1$.
}.
This vexing fact will be seen to be a consequence of our results in the next chapter. See especially section\ \ref{sec:collapse}, where we will argue that the eigenvectors of the finitely twisted chain in a given cyclicity sector at fixed $L$, $M$, $K$ all become collinear with a certain ``locked'' state. 
\section{\mathversion{bold}Collapse of the Bethe States in the Strong Twisting Limit}
\label{sec:QISM}

In this chapter, we will study the strong twisting $\varepsilon \rightarrow 0$ transition from the integrable three-state spin chain model twisted by three parameters $q_1, q_2, q_3$ to the general eclectic model with twisting parameters $\xi_1, \xi_2, \xi_3$ on the level of the ABA. In particular, we would like to understand the fate of the Bethe states \eqref{eq:ABA1} that diagonalize the transfer matrix \eqref{eq:transferev} for any $\varepsilon >0$, and investigate whether any useful information on the generalized eigenstates \eqref{eq:transfergenev} may be gathered from the limiting procedure.

In section \ref{sec:ABA} we will analyze the BAE and their solutions in terms of Bethe roots in the $\varepsilon \rightarrow 0$ limit. We uncover rather intricate and rich behavior, with, generically, {\it fractional} scaling behavior of the Bethe roots. Reassuringly, this will be shown to lead to the correct generalized eigenvalues \eqref{eq:transfergeneval} of the transfer matrix, thereby also explaining from the point of view of the ABA why all three-particle eigenstates of the eclectic chain are zero-energy states. However, despite the rich scaling behavior discovered in \ref{sec:ABA}, it appears that no further non-trivial  information on the structure of the Jordan blocks (i.e.\ their number and sizes) or the structure of the generalized eigenstates may be extracted from the ABA in the scaling limit: We find strong evidence in 
\ref{sec:collapse} that {\it all} Bethe states collapse to, in each cyclicity sector, a single ``locked state'' that happens to be the one true eigenstate of the largest Jordan block.

In the below, we will only study the transition of the ABA to the eclectic spin chain with ``generic'' scaled twist parameters $\xi_1$, $\xi_2$, $\xi_3$. The limit to the hypereclectic chain would have a different singularity structure and will not be analyzed. Therefore we may, without loss of generality,\footnote{
If these inequalities do not hold, one may suitably relabel the three pairs $(\phi_i,q_i)$. This is possible, since the twisted model is invariant under permutations of the three indices $i=1,2,3$. Note that the strongly twisted eclectic model \eqref{eq:stronglytwistedXXXfinal}, \eqref{eq:strongtwistedpermfinal} is also invariant under permutations of the $(\phi_i,\xi_i)$, while the hypereclectic model \eqref{eq:hypereclecticH}, \eqref{eq:hypereclecticP} is not. Incidentally, this is ultimately the reason why the hypereclectic model has a more complicated spectrum than the eclectic one, see chapter \ref{sec:results}.
}
assume that the number of 1's making up the spin chain states is larger or equal than the number of 2's, while the latter is larger or equal than the number of 3's. Denoting these three numbers by $L-M$, $M-K$ and $K$, respectively, this is represented by the following inequalities:
\begin{equation}
L-M\ge M-K\ge K\quad\Leftrightarrow\quad L\ge 2M-K \quad {\rm and} \quad  M\ge 2K.
\label{eq:LMKdomain}
\end{equation}
These inequalities are saturated in the case $L=3 K$, $M=2 K$ for arbitrary $K$. This clearly corresponds to the special case where the spin chain is made up from an equal number of $1$'s, $2$'s and $3$'s.
\subsection{Bethe Ansatz for the Eclectic Spin Chain}\label{sec:ABA}
In a rather routine if somewhat tedious fashion one may derive a nested BAE for the level-one Bethe roots $u_1, \ldots, u_M$ as well as the level-two Bethe roots $v_1, \ldots, v_K$ from the ABA \eqref{eq:ABA1} and \eqref{eq:ABA2} for the eigenstates. The procedure employs integrability by using the so-called {\it fundamental commutation relations} derived from the Yang-Baxter based \eqref{eq:RTT}, see e.g.\ \cite{Levkovich-Maslyuk:2016kfv}. The final result\footnote{
Surprisingly, we have not been able to explicitly find this result for general $q_1$, $q_2$, $q_3$ in any of the literature we know. However, we do not want to claim originality here, as the procedure is so standard.
}
for the system of BAE of the three-state model for general twisting parameters $q_1$, $q_2$, $q_3$ reads\footnote{
Actually, at finite twist it is not necessary to assume that the ``filling inequalities'' \eqref{eq:LMKdomain} hold. Furthermore, there are six distinct (but fully equivalent) systems of BAE, related to \eqref{eq:betheeqsu},\eqref{eq:betheeqsv} by the six permutations of the $q_1,q_2,q_3$ as well as the corresponding interpretation of $M$ and $K$. The reason is, that we may choose any of the three particles $1,2,3$ for the reference state of the first-level Bethe ansatz, and any of the two remaining particles of the reference state of the (inhomogeneous) second-level Bethe ansatz. However, the other five systems will behave differently in the strong twisting limit. Here we will only discuss the case \eqref{eq:LMKdomain}.
}
\begin{eqnarray} 
\label{eq:betheeqsu}
\left(\frac{u_m+1}{u_m}\right)^L&=&\frac{{q_3}^L}{({q_1}{q_2}{q_3})^{K}}
\prod_{\substack{j=1\\ j\neq m}}^M\frac{u_m-u_j+1}{u_m-u_j-1}\prod_{i=1}^K\frac{u_m-v_i-1}{u_m-v_i},
\quad m=1, \ldots M,
\\ 
\label{eq:betheeqsv}
1&=&
\frac{({q_2}{q_3})^L}{({q_1}{q_2}{q_3})^{M}}
\prod_{j=1}^M\frac{v_l-u_j+1}{v_l-u_j}\prod_{\substack{i=1\\ i\neq l}}^K\frac{v_l-v_i-1}{v_l-v_i+1},
\quad l=1, \ldots K,
\end{eqnarray}
while the eigenvalues of the transfer matrix \eqref{eq:transferev} are then expressed as
\begin{eqnarray}\label{eq:trtwist}
\tilde{T}_{(q_1,q_2,q_3)}(u)&=&\frac{q_2^K}{q_3^{M-K}} (u+1)^L\ \prod_{m=1}^M\frac{u-u_m-1}{u-u_m} \\
&+&
\frac{q_3^{L-M}}{q_1^{K}} u^L\ \prod_{m=1}^M\frac{u-u_m+1}{u-u_m}\
\prod_{l=1}^K\frac{u-v_l-1}{u-v_l}
+
\frac{q_1^{M-K}}{q_2^{L-M}} u^L\ \prod_{l=1}^K\frac{u-v_l+1}{u-v_l}. \nonumber
\end{eqnarray}
The shift operator eigenvalue \eqref{eq:shiftev} is extracted as 
\begin{equation}\label{eq:shiftevABA}  
\tilde{T}_{(q_1,q_2,q_3)}(0)=\omega_L^k=\frac{q_2^K}{q_3^{M-K}}\,\prod_{m=1}^M \frac{u_m+1}{u_m}\,,
\end{equation}
while the energy, i.e.\ the eigenvalue of the Hamiltonian \eqref{eq:energyev}, is given by
\begin{equation}\label{eq:energyevABA}
\tilde E_{(q_1,q_2,q_3)}=\frac{\frac{d}{d u} \tilde{T}_{(q_1,q_2,q_3)}(0)}{\tilde{T}_{(q_1,q_2,q_3)}(0)}=
L+\sum_{m=1}^M\frac{1}{u_m (u_m+1)}=L+\sum_{m=1}^M\left(\frac{1}{u_m} - \frac{1}{u_m+1}\right).
\end{equation}
We have omitted the index $j$, labelling the eigenstates, on $\tilde{T}^j_{(q_1,q_2,q_3)}(u)$ and  $E^j_{(q_1,q_2,q_3)}$ in \eqref{eq:trtwist}, \eqref{eq:shiftevABA}, \eqref{eq:energyevABA}. Of course the Bethe roots strongly depend on $j$.

Let us now understand what happens to the BAE and their solutions in the strong twisting limit $\varepsilon \rightarrow 0$, where one replaces according to \eqref{eq:epsilon} (in the case $(+,+,+)$ with $\xi_j^+=\xi_j$) the $q_j$'s by
\begin{equation}\label{eq:scaledtwist}
q_j=\frac{\xi_j}{\varepsilon}\,\quad j=1,2,3\,
\end{equation}
in terms of new, finite deformation parameters $\xi_j$.
Inserting this into \eqref{eq:betheeqsu},\eqref{eq:betheeqsv} one obtains
\begin{eqnarray}
\left(\frac{u_m+1}{u_m}\right)^L&=&\varepsilon^{3K-L}\ \frac{{\xi_3}^L}{{\xi}^{K}}\
\prod_{\substack{j=1\\ j\neq m}}^M\frac{u_m-u_j+1}{u_m-u_j-1}\prod_{i=1}^K\frac{u_m-v_i-1}{u_m-v_i}\,,
\quad m=1, \ldots M,
\label{eq:baeu}
\\
1&=&
{\varepsilon}^{3M-2L}\ \frac{\xi^{L-M}}{\xi_1^L}\
\prod_{j=1}^M\frac{v_l-u_j+1}{v_l-u_j}\prod_{\substack{i=1\\ i\neq l}}^K\frac{v_l-v_i-1}{v_l-v_i+1}\,,
\quad l=1, \ldots K,
\label{eq:baev}
\end{eqnarray}
where we have introduced for conciseness of notation
\begin{equation}
\xi := \xi_1\xi_2\xi_3\,.
\end{equation}
Interestingly, for the special case
\begin{equation}\label{eq:specialfilling}
L=3K\,, \quad M=2K\,, \quad {\rm any}\,\,\, K\,,
\end{equation}
the powers of $\varepsilon$ drop out, and the BAE, and hence their solutions, are identical to the ones of the twisted model {\it before} taking the scaling limit, \eqref{eq:betheeqsu},\eqref{eq:betheeqsv}, with the original $q_j$'s simply replaced by the $\xi_j$'s. See \ref{sec:3K2KK} below. In all other cases the powers of $\varepsilon$ lead to inconsistencies, unless the Bethe roots exhibit a suitably singular behavior so as to compensate these powers. Finding the correct scaling of the roots can be quite tricky, especially for the level-two roots $\{v_m\}$. We will now present a near-complete classification (see \ref{sec:remainingLMK} for the still unclear cases) of the possible scaling behaviors, and their consequences for the ``observables'' \eqref{eq:trtwist}, \eqref{eq:shiftevABA}, \eqref{eq:energyevABA}.

\subsubsection{\mathversion{bold}$K=0$ Sector}
\label{sec:K=0}

This corresponds to an $n=2$ chiral XY model, where only the $u_m$ Bethe roots are kept.
It is not necessary to assume the filling condition $L\geq 2M$. The nested BAE \eqref{eq:baeu} turn into the unnested set
\begin{eqnarray}
\left(\frac{u_m+1}{u_m}\right)^L=\frac{{\xi_3}^L}{\varepsilon^{L}}\
\prod_{\substack{j=1\\ j\neq m}}^M\frac{u_m-u_j+1}{u_m-u_j-1}\,,
\quad m=1, \ldots M.
\label{eq:baeK0}
\end{eqnarray}
It is rather obvious, and discussed in considerable detail in \cite{Ipsen:2018fmu}, that the Bethe roots should scale as $u_m=\varepsilon\, u^-_m$ in the $\varepsilon\to 0$ limit. The BAE \eqref{eq:baeK0} then reduce to 
\begin{equation}\label{eq:baeK=0}
(\xi_3\, u^-_m)^L=1\,,
\quad m=1, \ldots M.
\end{equation}
The solutions $\{u_m\}$ are immediately seen to be given in terms of subsets of the $L$ $L$-th roots of unity. It is easy to demonstrate completeness of states of these equations, in line with the fact that the $K=0$ sector is completely diagonalizable, despite the strongly twisted two-state model's nonhermiticity. No non-trivial (i.e.\ of size larger than one) Jordan blocks appear. The transfer matrix eigenvalue \eqref{eq:trtwist} turns into
\begin{equation}\label{eq:trtwistK=0}
\hat{T}_{(\xi_1,\xi_2,\xi_3)}(u)=\frac{(\xi_3\, u)^L+(-1)^M}{\displaystyle{\prod_{m=1}^M }\xi_3\,(u-u_m^-)}+\delta_{L,M}\,(\xi_1\,u)^L\,.
\end{equation}
The shift operator eigenvalue \eqref{eq:shiftevABA} is then given by
\begin{equation}\label{eq:shiftevABAK=0}  
\hat{T}_{(\xi_1,\xi_2,\xi_3)}(0)=\omega_L^k=\prod_{m=1}^M \frac{1}{\xi_3\, u_m^-}\,,
\end{equation}
while the energy, i.e.\ the eigenvalue of the Hamiltonian \eqref{eq:energyevABA}, is elegantly given by
\begin{equation}\label{eq:energyevABAK=0}
\hat E_{(\xi_1,\xi_2,\xi_3)}=
\sum_{m=1}^M \frac{1}{u_m^-}\,.
\end{equation}

\subsubsection{\mathversion{bold}$L=3 K$, $M=2 K$ Sector}\label{sec:3K2KK}

In the $\varepsilon\to 0$ limit, the BAE \eqref{eq:baeu} and \eqref{eq:baev} look singular except for the special case
\begin{equation}
L=3K\,, \quad M=2K\,,
\end{equation}
where the $\varepsilon$-dependence simply drops out.
In this curious case, for the strongly twisted model the {\it same} BAE as for the original generic twisted model are satisfied, except that
the three twist parameters $q_i$ are replaced by $\xi_i$ in \eqref{eq:baeu}, \eqref{eq:baev}. This means that the Bethe roots of a given state do not change at all as one approaches the strong twist limit! However, the effects of the limit are still seen on the level of the expressions for the ``observables''. In particular,
since we also need to replace $u \rightarrow \epsilon\, u$ along with \eqref{eq:scaledtwist},
we find that the expression for the transfer matrix eigenvalue \eqref{eq:trtwist} becomes a $u$-independent constant, and thus identical to the shift operator eigenvalue \eqref{eq:shiftevABA}:
\begin{equation}\label{eq:transferspecial}  
\hat{T}_{(\xi_1,\xi_2,\xi_3)}(u)=\hat{T}_{(\xi_1,\xi_2,\xi_3)}(0)=\omega_L^k=\left(\frac{\xi_2}{\xi_3}\right)^K\,\prod_{m=1}^M \frac{u_m+1}{u_m}\,,
\end{equation}
while the energy formula \eqref{eq:energyevABA} formally yields
\begin{equation}\label{eq:energyspecial}
\hat E_{(\xi_1,\xi_2,\xi_3)}=
0\,.
\end{equation}
This is rather remarkable: Even though the BAE keep their full complexity in the strong twisting limit, and thus their solutions their full intricacy, the effect on the energy is irrelevant: Since the roots stay finite, and the energy formula \eqref{eq:energyevABA} has to be multiplied by $\varepsilon$, one trivially obtains zero in the limit. Likewise, even the result for the transfer matrix eigenvalue does not really depend on the details of the Bethe roots, as the last of the identities in \eqref{eq:transferspecial} is a simple, direct consequence of the BAE.\footnote{To show this, using a standard trick, first take the product of the $M$ equations \eqref{eq:baeu}, and then simplify the resulting r.h.s.\ with the help of the product of the $K$ equations \eqref{eq:baev}.
}
In conclusion, for $L=3 K$, $M=2 K$ at infinite twist there is no point in solving the BAE! The only two things one needs to know are that they stay non-singular (i.e.\ $\neq 0,-1$) and finite in the limit, and that they yield a complete set of states for this sector. On the other hand, the sector turns out to become non-diagonalizable in the scaling limit, with an abundance of Jordan blocks appearing. Unfortunately, this already gives a strong hint that the conventional Bethe ansatz is useless as regards the explanation of the model's Jordan block structure. We will soon see better why this is the case, cf.\ section \ref{sec:collapse}. However, for completeness, let us first understand the scaling of the Bethe roots away from the special, symmetric filling conditions of the $L=3 K$, $M=2 K$ sector.

\subsubsection{\mathversion{bold}$M=2$, $K=1$ sector}\label{sec:L21}

It is illuminating to first analyze the two-excitation case $M=2, K=1$ with general $L>3$ in some detail.\footnote{For $L=3$ we are back to the situation of section \ref{sec:3K2KK}.
}
From \eqref{eq:baeu}, \eqref{eq:baev} we then have a set of three BAE, which explicitly read
\begin{eqnarray}
\left(\frac{u_1+1}{u_1}\right)^L&=&\varepsilon^{3-L}\ \frac{\xi_3^{L}}{\xi}\
\left(\frac{u_1-u_2+1}{u_1-u_2-1}\right)\left(\frac{u_1-v_1-1}{u_1-v_1}\right),
\label{eq:equ1}
\\
\left(\frac{u_2+1}{u_2}\right)^L&=&\varepsilon^{3-L}\ \frac{\xi_3^{L}}{\xi}\
\left(\frac{u_2-u_1+1}{u_2-u_1-1}\right)\left(\frac{u_2-v_1-1}{u_2-v_1}\right),
\label{eq:equ2}
\\
1&=&\varepsilon^{6-2L}\ \frac{\xi^{L-2}}{\xi_1^{L}}\
\left(\frac{v_1-u_1+1}{v_1-u_1}\right)\left(\frac{v_1-u_2+1}{v_1-u_2}\right).
\label{eq:eqv1}
\end{eqnarray}
From these, one notices that the Bethe roots should have the following specific scaling form in the $\varepsilon\to 0$ limit:
\begin{equation}
u_1=\varepsilon^{\alpha}\, u^-,\quad
u_2=-1+\varepsilon^{\alpha}\, u^+,\quad
v_1=u_2-1+\varepsilon^\gamma\, {\hat v}\,,
\label{eq:scale21}
\end{equation}
where $\alpha$, $\gamma$ are positive exponents and the ``scaled'' Bethe roots $u^{\pm}$ and $\hat v$ are finite. Except for the general powers of $\varepsilon$, the situation is as described in \cite{Ipsen:2018fmu}: One of the excitations is a ``right-mover'', the other a ``left-mover''. However, here we do not assume $\alpha=1$.
Then, inserting this ansatz into the BAE, one finds that the powers are generically\footnote{This scaling behavior holds for ``generic'' values of the twist parameters $\xi_i$. It may be modified by a suitable finetuning of the latter. Furthermore, note that for $L=3$ \eqref{eq:scale21} predicts $\alpha=\gamma=0$, which is consistent with section \ref{sec:3K2KK}: The roots in \eqref{eq:scale21} stay finite.
}
given by
\begin{equation}
\alpha=\frac{L-3}{L-1},\quad \gamma=2L-6\,.
\label{eq:exp21}
\end{equation}
It is interesting to note that the exponent $\alpha$ is a fractional number. This is quite unexpected, and has not been noticed before in the existing literature on the subject.
The scaled Bethe roots then satisfy at generic $\xi_i$ the simplified BAE
\begin{equation}\label{eq:BAE12}
(u^-)^L=\frac{\xi}{\xi_3^L}\, (u^- - u^+)\,,\quad
(-u^+)^{L}
=\frac{\xi}{\xi_2^L}\,(u^- - u^+)
\,,\quad
{\hat v}=-\frac{2\xi_1^L}{\xi^{L-2}}\,.
\end{equation}
The solutions of these equations can be found explicitly in terms of the $L$-th roots and the $(L-1)$-th roots of unity. One easily shows completeness of all $L (L-1)$ states of this sector.
These solutions show a somewhat intricate dependence on the scaled twist parameters $\xi_i$. We will not exhibit them here, as they are not really needed, in close analogy with the previous section \ref{sec:3K2KK}. As we can see from \eqref{eq:exp21}, the scaling exponent $\alpha$ is bounded by $\frac{1}{3} \leq \alpha < 1$. Hence, the energy formula \eqref{eq:energyevABA} yields again in the $\varepsilon \to 0$ limit
\begin{equation}\label{eq:energy21}
\hat E_{(\xi_1,\xi_2,\xi_3)}=
0\,,
\end{equation}
and the detailed values of the scaled Bethe roots $u^+, u^-$ are wiped out. The same is again true for the transfer matrix eigenvalue \eqref{eq:trtwist}, which once more becomes a $u$-independent constant, and thence identical to the shift operator eigenvalue \eqref{eq:shiftevABA}:
\begin{equation}\label{eq:transfer12}  
\hat{T}_{(\xi_1,\xi_2,\xi_3)}(u)=\hat{T}_{(\xi_1,\xi_2,\xi_3)}(0)=\omega_L^k=-\frac{\xi_2}{\xi_3}\,\frac{u^+}{u^-}\,.
\end{equation}
While this expression appears to depend on the detailed solution for $u^+$,$u^-$, one sees from 
\eqref{eq:BAE12} that the constant must indeed be equal to the eigenvalue $\omega_L^k$ of the shift operator.
Vexingly, while we are able to find the scaling limit of the sector's BAE, and could even solve them exactly, we see no hint, in full analogy with section \ref{sec:3K2KK}, on how to derive the sector's Jordan block structure. The latter is actually quite simple here, and given by a single Jordan block of size $L\!-\!1$ in each of the $L$ cyclicity sectors, cf.\ chapter \ref{sec:results}.

\subsubsection{\mathversion{bold} Generic $(L,M, K)$ Sector with $L>3(M-K)$}
\label{sec:LMK}

Let us generalize the analysis of the $M=2,K=1$ case of \ref{sec:L21}, and assume that the level-one roots $u_m$ 
($m=1,\ldots,M$) can be split into classes (I) and (II) of right/left-movers:
\begin{eqnarray}
&({\rm I})&\qquad \quad u_j=\varepsilon^{\alpha}\, u^-_j\,,\qquad \qquad j=1,\cdots, M' := M-K\,,
\label{eq:scaleu1}\\
&({\rm II})&\quad \,\, u_{l+M'}=-1+\varepsilon^{\beta}\, u^+_l\,,\quad \,\, l=1,\cdots, K\,.
\label{eq:scaleu2}
\end{eqnarray}
%
For each class (II) root, we associate a level-two Bethe root $v_k$, which we term
class (III). For its scaling behavior we make the ansatz
\begin{equation}
({\rm III})\quad v_l=u_{l+M'}-1+\varepsilon^{\gamma}\, {\hat v}_l=-2+\varepsilon^{\beta}\, u^+_l+\varepsilon^{\gamma}\, {\hat v}_l,\qquad l=1,\cdots, K\,.
\label{eq:scalev}
\end{equation}
Our main assumption is that the Bethe roots in each class become degenerate in the leading order of $\varepsilon\to 0$, with small deviations appearing
at subleading order with positive powers $\alpha,\beta,\gamma$ of  $\varepsilon$.
While $\alpha=\beta$ for the case of $M=2, K=1$ in \eqref{eq:scale21}, we need to assume 
$\alpha\neq\beta$ for the general case.
Inserting this ansatz into the system of BAE \eqref{eq:baeu}, \eqref{eq:baev}, the class (I) roots from \eqref{eq:baeu} satisfy
\begin{eqnarray}
(u^-_j)^L&=&(-1)^{M'-1+K}\, \frac{\xi^K}{\xi_3^L}\ \prod_{l=1}^Ku^+_l,
\label{sbae1}\\
\alpha\, L&=&L-3K+{\rm min}(\alpha,\beta)\, K.
\label{eq:cond1}
\end{eqnarray}
Similarly, \eqref{eq:baeu} for the class (II) roots gives
\begin{eqnarray}
(u^+_l)^{L-M'}&=&(-1)^{L-M}\, \frac{\xi_3^L}{2^{M'}\xi^K}\ {\hat v}_l\prod_{\substack{i=1\\ i\neq l}}^K
(u^+_l-u^+_i),
\label{eq:sbae2}\\
\beta\, L&=&L-3M'+{\rm min}(\alpha,\beta)\, M'.
\label{eq:cond2}
\end{eqnarray}
Finally, (\ref{eq:baev}) for the class (III) roots along with \eqref{eq:gambet} yields
\begin{eqnarray}
1&=&-\frac{\xi^{L-M}}{2^{M'}\xi_1^L}\ {\hat v}_l\prod_{\substack{i=1\\ i\neq l}}^K
(u^+_l-u^+_i),
\label{sbae3}\\
2L-3M&=&\gamma+\beta (K-1).
\label{eq:cond3}
\end{eqnarray}
We have assumed 
\begin{equation}
\gamma>\beta
\label{eq:gambet}
\end{equation}
in deriving \eqref{eq:cond2} and \eqref{eq:cond3}, which will be justified shortly.
From the difference of \eqref{eq:cond1} and \eqref{eq:cond2}, one deduces 
$\alpha\ge\beta$ if $M'\ge K$, which is consistent with \eqref{eq:LMKdomain}. 
Furthermore, solving
\eqref{eq:cond1}, \eqref{eq:cond2}, and \eqref{eq:cond3}, one finds for the scaling exponents
\begin{eqnarray}
\alpha&=&\frac{L-(M+K)}{L-(M-K)}\,,\label{eq:alpha}\\
\beta&=&\frac{L-3(M-K)}{L-(M-K)}\,,\label{eq:beta}\\
\gamma&=&2L-3M-\frac{L-3(M-K)}{L-(M-K)}\ (K-1).\label{eq:gamma}
\end{eqnarray}
Note that the exponents \eqref{eq:exp21} for the special cases $K=0$ in \ref{sec:K=0} and $M=2, K=1$ in section \ref{sec:L21} are consistent with these results.
For the smallest exponent $\beta$ to be positive, the inequality 
\begin{equation}
L>3\,(M-K)
\label{eq:domain}
\end{equation}
should be satisfied.
It is consistent\footnote{This limit does not apply to $K=0$, cf.\ section \ref{sec:K=0}, where the level-two Bethe roots are not needed.
}
with \eqref{eq:LMKdomain}.
The assumption \eqref{eq:gambet} is valid, since

\begin{equation}
\gamma-\beta=\frac{(2L-3M')(L-M-K)}{L-M'}>0\,,
\end{equation}
as long as \eqref{eq:LMKdomain} and \eqref{eq:domain} are satisfied.
Now we can even explicitly solve for the scaled Bethe roots in \eqref{eq:scaleu1}, \eqref{eq:scaleu2}, \eqref{eq:scalev}. By plugging (\ref{sbae3}) into \eqref{eq:sbae2}, we get
\begin{equation}
(u^+_l)^{L-M'}=(-1)^{L-M+1}\frac{(\xi_1\xi_3)^L}{\xi^{L-M'}}\quad
\Rightarrow
\quad u^+_l=\frac{1}{\xi}(\xi_1\xi_3)^{\frac{L}{L-M'}}\omega_{L-M'}^{n_l}\,,
\label{eq:solup}
\end{equation}
recalling the notation $\omega_L=\exp(2\pi i/L)$ in \eqref{eq:rootfunity}.
For each root $u^{+}_l$ with $l=1,\cdots,K$, the $K$ integer exponents 
$n_l$ should be distinctly chosen among the set $\{1,\cdots,L-M'\}$.
Once the $u^{+}_l$ are determined, the Bethe roots ${\hat v}_l$ are uniquely fixed by
\eqref{sbae3}. In the same way, the Bethe roots $u^-_j$ can be determined from \eqref{sbae1}:
\begin{equation}
u^-_j=\left[(-1)^{M'-1+K}\frac{\xi^K}{\xi_3^L}\ \prod_{l=1}^Ku^+_l\right]^{1/L}\ \omega_L^{i_j},
\label{eq:solum}
\end{equation}
where the $M'$ integer exponents 
$i_j$ should be chosen without repetition from the set $\{1,\cdots,L\}$.

Let us quickly check completeness of states. The number of solutions of the BAE may easily be computed using the standard combinatorial identity
\begin{equation}
\left(\begin{array}{c} L-M'\\ K\end{array}\right)
\left(\begin{array}{c} L\\ M'\end{array}\right)=\frac{L!}{(L-M)!(M-K)!K!}\,,
\label{eq:totalconfig}
\end{equation}
where we recall $M'=M-K$.
Notice that this is exactly the dimension of the transfer matrix in the $(L,M,K)$ sector,
and hence the number of its generalized eigenvalues. This shows that the above set of solutions of the BAE in the scaling limit is indeed complete.

Using our scaling solution, we may now again formally compute the transfer matrix eigenvalue \eqref{eq:trtwist}, which once more becomes a $u$-independent constant, entirely due to the first of the three terms in \eqref{eq:trtwist}:
\begin{equation}\label{eq:transferLMK}
\hat{T}_{(\xi_1,\xi_2,\xi_3)}(u)=\frac{(-\xi_2)^K}{\xi_3^{M-K}} \varepsilon^{M-2K}
\varepsilon^{\beta K-\alpha(M-K)}\ 
\prod_{l=1}^Ku^+_l\Bigg/ \prod_{j=1}^{M'}u^-_j\,.
\end{equation}
Here we made use of the fact that $\alpha,\beta<1$. Note that the power of $\varepsilon$ 
cancels with the scaling exponents $\alpha$, $\beta$ from \eqref{eq:alpha},\eqref{eq:beta}. 
Also, using the explicit solutions \eqref{eq:solup} and \eqref{eq:solum}, we find 
the eigenvalues are given by 
\begin{equation}
\hat{T}_{(\xi_1,\xi_2,\xi_3)}(u)=\omega_L^k\,,\quad
k=\sum_{l=1}^Kn_l-\sum_{j=1}^{M-K}i_j+\frac{1}{2}(M-K)(M-K-1)+\frac{1}{2}K(K-1)\,.
\label{eq:explicitEV}
\end{equation}
Since the exponent $k$ is an integer, the eigenvalues are indeed the $L$-th roots of unity, as expected.
Note that $k$ explicitly depends on the quantum numbers introduced to label the solutions for the Bethe roots. 
In appendix \ref{app:polya}, we show that this result is consistent with the famous P\'olya enumeration theorem for cyclic states.

The expression \eqref{eq:transferLMK} actually also comprises the results for the earlier cases
\eqref{eq:transferspecial} (where $\alpha=\beta=0$) and \eqref{eq:transfer12}. Clearly, we find from \eqref{eq:transferLMK} that, once again, the (generalized) energy eigenvalues, given by logarithmic derivatives of the transfer matrix eigenvalues at $u=0$, are all found to be
\begin{equation}\label{eq:energyLMK}
\hat E_{(\xi_1,\xi_2,\xi_3)}=
0\,.
\end{equation}

Curiously, while we were able to understand the rather involved scaling limit of this large sector's BAE, and could even solve them exactly, we see no hint, in full analogy with the more special results of sections \ref{sec:3K2KK}, \ref{sec:L21}, on how to derive the sector's Jordan block structure. As opposed to section \ref{sec:L21}, the latter is actually very complicated here, and we currently do not understand its general pattern, cf.\ chapter \ref{sec:results}.

\subsubsection{\mathversion{bold} Remaining Values for $(L,M, K)$}
\label{sec:remainingLMK}
The results of the previous sections do not fully cover all the regions in \eqref{eq:LMKdomain}.
They are not valid for the (relatively small) ``window'' where \eqref{eq:LMKdomain} holds 
but \eqref{eq:domain} fails, namely,
\begin{equation}
3(M-K)\ge L\ge 2M-K\,.
\label{eq:excludeddomain}
\end{equation}
The special case $L=3K, M=2K$ in \ref{sec:3K2KK} actually happens to be the boundary case of this domain, where the above inequalities turn into equalities. As we argued in \ref{sec:3K2KK}, there the Bethe roots do not scale at all and remain finite.
We therefore suspect that the Bethe roots of the missing region exhibit some mixed scaling behavior, where some of the roots stay finite, while others turn to zero or $-1$. While it would be technically pleasing to untangle this, we are convinced that it will not really bring any new insights: The (generalized) eigenvalue of the transfer matrix will presumably still be given by the r.h.s.\ of \eqref{eq:transferLMK} as a $u$-independent constant equal to the eigenvalue of the shift operator. The (generalized) energy eigenvalue will once more be zero. And no information on the Jordan block structure will be contained in the details of the Bethe roots, unfortunately. For these reasons we will not pursue the further study of this region.
\subsection{\mathversion{bold}Collapse of the Bethe States}\label{sec:collapse}

As explained in section \ref{sec:failure}, in the framework of the standard quantum inverse scattering method, the eigenstates are constructed by acting with certain creation operators on the level-one and level-two reference states as in \eqref{eq:ABA1} and \eqref{eq:ABA2}, respectively.
Each set of Bethe roots satisfying the BAE \eqref{eq:betheeqsu}, \eqref{eq:betheeqsv}
defines a corresponding eigenstate. 
This leads to an explicit construction of these eigenstates, in principle.
Now we want to study the eigenstates of the strongly twisted model by taking the $\varepsilon \to 0$ limit of these Bethe states. 
We have already shown that the Bethe roots become highly degenerate in this limit,
$u_j=0, -1$ and $v_l=-2$, in the leading order as summarized in
\eqref{eq:scaleu1}, \eqref{eq:scaleu2}, \eqref{eq:scalev} with small corrections suppressed by
various positive powers of $\varepsilon$.
Hence, it is natural to expect  that the eigenstates will also become degenerate. 

An important initial observation from \eqref{eq:commutation} is the commutation relation\footnote{We will omit the subscripts $(q_1,q_2,q_3)$ and $(\xi_1,\xi_2,\xi_3)$ for simplicity in this section.}
\begin{equation}
\left[\mathbf{\tilde T}(u),\mathbf{\tilde T}(0)\right]=0,
\label{eq:commuTT}
\end{equation}
which imposes that any eigenstate $\vert\psi\rangle$ of $\mathbf{\tilde T}(u)$ should also be one of $\mathbf{\tilde T}(0)$, the shift operator 
$\mathbf{U}$ in \eqref{eq:shiftev}.
Since $\mathbf{\tilde T}(0)^L=\mathbb{I}$, the Bethe states should satisfy
\begin{equation}
\mathbf{\tilde T}(u)\vert\psi_{\Lambda,k}\rangle=\Lambda\vert\psi_{\Lambda,k}\rangle,\qquad
\mathbf{\tilde T}(0)\vert\psi_{\Lambda,k}\rangle=\omega_L^k\,\vert\psi_{\Lambda,k}\rangle, \quad k=0,\cdots,L-1.
\label{eq:Taction}
\end{equation}
For a given configuration $\mathbf{f}$, we define a shifted configuration $\pi(\mathbf{f})$ 
\begin{equation}
\mathbf{f}=(n_1,n_2,\cdots,n_{L-1},n_{L}),\quad n_j=\{1,2,3\}\quad : \Rightarrow\quad
\pi(\mathbf{f})=(n_{L},n_1,n_2,\cdots,n_{L-1}).
\label{eq:shiftedconf}
\end{equation}
Then, the action of the shift operator on a state gives
\begin{equation}
\mathbf{\tilde T}(0)\vert \mathbf{f}\rangle=\vert \pi(\mathbf{f})\rangle.
\label{eq:shift2}
\end{equation}
In terms of these notations, we can find eigenstates\footnote{Note that it may happen that for some $| \mathbf{f} \rangle$ and $k$ one might have $| \mathbf{f}_k \rangle=0$, in which case $| \mathbf{f}_k \rangle$ is not an eigenstate, of course. A trivial example is 
$| \mathbf{f} \rangle=|1 1 \ldots 1 \rangle$ for $k \neq 0$. A second example would be 
$| \mathbf{f} \rangle=|1 2 1 2 \rangle$, where $| \mathbf{f}_1 \rangle=| \mathbf{f}_3 \rangle=0$, while $| \mathbf{f}_0 \rangle$ and $| \mathbf{f}_2 \rangle$ are eigenstates with eigenvalues $1$, $-1$, respectively. 
This is related to P\'olya counting, see appendix \ref{app:polya}.
}
of the shift operator for a given configuration $\mathbf{f}$ as follows:
\begin{equation}\label{eq:cyclicbasis}
\vert\mathbf{f}_k\rangle=\sum_{\ell=1}^L(\omega_L^k)^{-\ell}\vert \pi^\ell(\mathbf{f})\rangle,
\end{equation}
where $\pi^\ell$ means a shift by $\ell$ steps.
The case of $k=0$ corresponds to the cyclic states.

Now the simultaneous eigenstate $\vert\psi_{\Lambda,j}\rangle$ in \eqref{eq:shift2} can be 
expressed as a linear combination of the eigenstates $\vert\mathbf{f}_k\rangle$ of the shift operator:
\begin{equation}\label{eq:shiftevbasis}
\vert\psi_{\Lambda,k}\rangle=\sum_{\mathbf{f}} a({\mathbf{f}})\vert\mathbf{f}_k\rangle,
\end{equation}
where the sum is over all possible configurations that are not related by shifts.
The Bethe states in \eqref{eq:ABA1} constructed by the creation operators and associated Bethe roots, 
in principle, determine the coefficients $a_j({\mathbf{f}})$, 
although actual computations for generic $(L,M,K)$ can be very complicated.
However, in the strong twisting limit $\varepsilon\to 0$, it is possible to find a configuration 
$\mathbf{f}$ that makes the coefficient 
$a({\mathbf{f}})$ most dominant in powers of $\varepsilon$ by analyzing the creation operators and their actions on the reference state. 

The creation operators $\mathcal{B}_{1}=\mathbf{\tilde{M}}_{12}$ and 
$\mathcal{B}_{2}=\mathbf{\tilde{M}}_{13}$ in the Bethe state \eqref{eq:ABA1}
are composed of the $\mathbf{\tilde{R}}$ matrices as defined in \eqref{eq:monodromy}.
Our strategy is to express the Bethe states graphically by two-dimensional square lattices, 
where each vertex corresponds to Boltzmann weights defined by the $\mathbf{\tilde{R}}$ matrix.
In the $\varepsilon\to 0$ limit, we can determine how these Boltzmann weights scale with $\varepsilon$, thereby yielding the leading configurations for each eigenstate.

The $\mathbf{\tilde{R}}^{a,n}(u)$ in \eqref{eq:Rmat} may be written as
\begin{equation}\label{eq:Rmatnew}
\mathbf{\tilde{R}}^{a,\ell}(u)=(u+1)\sum_{i=1}^3e^{a}_{ii}\otimes e^\ell_{ii}
+u\sum_{\substack{{\rm even}\\ (i,j,k)}}\left(\frac{\xi_j}{\varepsilon}\right)e^{a}_{ii}\otimes e^\ell_{kk}
+u\sum_{\substack{{\rm odd}\\ (i,j,k)}}\left(\frac{\varepsilon}{\xi_j}\right)e^{a}_{ii}\otimes e^\ell_{kk}
+\sum_{i\neq j}e^{a}_{ij}\otimes e^n_{ji},
\end{equation}
where ``even/odd'' in the sum mean even/odd permutations of $(ijk)$.
Here we use the concise notation $e^a_{ij}$, $e^\ell_{ij}$ for $3\times 3$ matrices acting on auxiliary space (a) or the $\ell$-th quantum space, respectively, 
whose elements are given by $(e_{ij})_{ab}=\delta_{ai}\delta_{bj}$.
We should evaluate $\mathbf{\tilde{R}}^{a,\ell}(u_j)$ with a Bethe root $u_j$ since the arguments of the creation operators in the Bethe state \eqref{eq:ABA1} are Bethe roots, the solutions of 
the BAE \eqref{eq:betheeqsu}, \eqref{eq:betheeqsv}.
For the generic $(L,M,K)$ sector with $L>3(M-K)$, which corresponds to 
sections \ref{sec:L21} and \ref{sec:LMK}, 
the Bethe roots are scaling as either class (I) 
\eqref{eq:scaleu1} or class (II) \eqref{eq:scaleu2} in the strong twisting limit.

\begin{figure}[t]
\setlength{\unitlength}{.1cm}
\begin{picture}(0,20)
\thicklines
\put(10,10){\line(1,0){20}}
\put(20,0){\line(0,1){20}}
\put(10,11){$i$}
\put(21,1){$k$}
\put(28,11){$i$}
\put(21,18){$k$}
\put(45,10){\line(1,0){20}}
\put(55,0){\line(0,1){20}}
\put(45,11){$i$}
\put(56,1){$j$}
\put(63,11){$j$}
\put(56,18){$i$}
\put(80,10){\line(1,0){20}}
\put(90,0){\line(0,1){20}}
\put(80,11){$i$}
\put(91,1){$i$}
\put(98,11){$i$}
\put(91,18){$i$}
\put(115,10){\line(1,0){20}}
\put(125,0){\line(0,1){20}}
\put(115,11){$i$}
\put(126,1){$k$}
\put(133,11){$i$}
\put(126,18){$k$}
\put(17,-7){$\frac{\xi_j u^-}{\varepsilon^{1-\alpha}}$}
\put(54,-7){$1$}
\put(89,-7){$1$}
\put(118,-7){$\frac{\varepsilon^{1+\alpha}u^-}{\xi_j}$}
\put(10,25){(a) even $(ijk)$}
\put(47,25){(b) $i\ne j$}
\put(87,25){(c)}
\put(116,25){(d) odd $(ijk)$}
\end{picture}
\vskip 1.cm
\caption{The Boltzmann weights of $\mathbf{\tilde{R}}^{a,\ell}$ for Bethe roots of the class (I)}
\label{Fig:BoltzmannWeight1}
\end{figure}
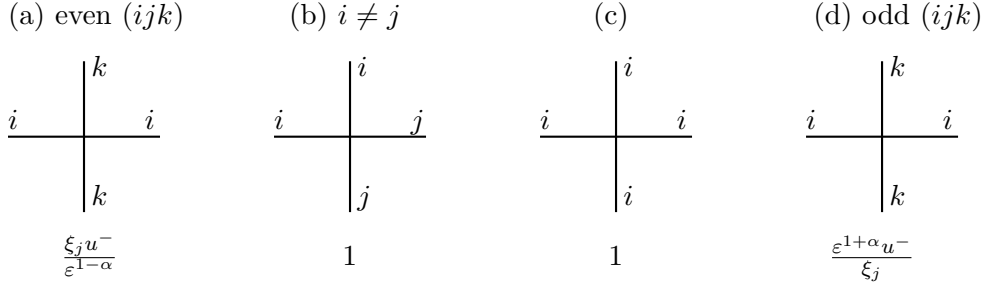

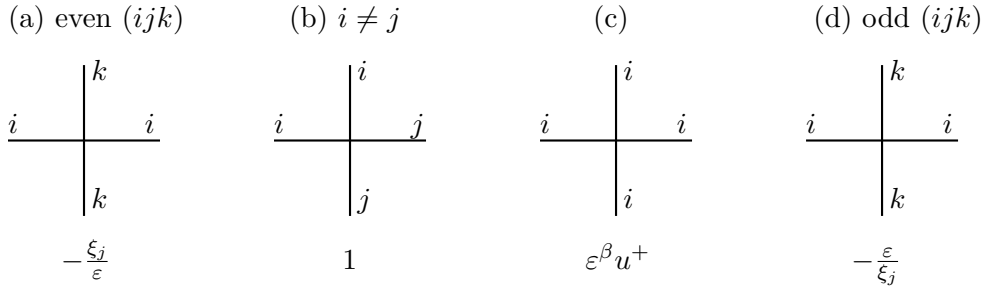
\begin{figure}[b]
\setlength{\unitlength}{.1cm}
\begin{picture}(0,25)
\thicklines
\put(10,10){\line(1,0){20}}
\put(20,0){\line(0,1){20}}
\put(10,11){$i$}
\put(21,1){$k$}
\put(28,11){$i$}
\put(21,18){$k$}
\put(45,10){\line(1,0){20}}
\put(55,0){\line(0,1){20}}
\put(45,11){$i$}
\put(56,1){$j$}
\put(63,11){$j$}
\put(56,18){$i$}
\put(80,10){\line(1,0){20}}
\put(90,0){\line(0,1){20}}
\put(80,11){$i$}
\put(91,1){$i$}
\put(98,11){$i$}
\put(91,18){$i$}
\put(115,10){\line(1,0){20}}
\put(125,0){\line(0,1){20}}
\put(115,11){$i$}
\put(126,1){$k$}
\put(133,11){$i$}
\put(126,18){$k$}
\put(17,-7){$-\frac{\xi_j}{\varepsilon}$}
\put(54,-7){$1$}
\put(86,-7){$\varepsilon^\beta u^+$}
\put(121,-7){$-\frac{\varepsilon}{\xi_j}$}
\put(10,25){(a) even $(ijk)$}
\put(47,25){(b) $i\ne j$}
\put(87,25){(c)}
\put(116,25){(d) odd $(ijk)$}
\end{picture}
\vskip 1.cm
\caption{The Boltzmann weights of $\mathbf{\tilde{R}}^{a,\ell}$ for Bethe roots of the class (II)}
\label{Fig:BoltzmannWeight2}
\end{figure}

For the Bethe roots $u^{\rm I}$ of class (I), $\mathbf{\tilde{R}}^{a,\ell}$ in 
\eqref{eq:Rmatnew} can be expanded as
\begin{equation}\label{eq:RmatnewI}
\mathbf{\tilde{R}}^{a,\ell}(u^{\rm I})=\frac{u^-}{\varepsilon^{1-\alpha}}\sum_{\substack{{\rm even}\\ (i,j,k)}}\xi_j e^{a}_{ii}\otimes e^\ell_{kk}
+\sum_{i\neq j}e^{a}_{ij}\otimes e^\ell_{ji}+\sum_{i=1}^3e^{a}_{ii}\otimes e^\ell_{ii}
+u^-\varepsilon^{1+\alpha}\sum_{\substack{{\rm odd}\\ (i,j,k)}} \frac{1}{\xi_j} e^{a}_{ii}\otimes e^\ell_{kk},
\end{equation}
in increasing powers of $\varepsilon$.
Similarly, $\mathbf{\tilde{R}}^{a,\ell}$ with the Bethe root $u^{\rm II}$ of class (II) becomes
\begin{equation}\label{eq:RmatnewII}
\mathbf{\tilde{R}}^{a,\ell}(u^{\rm II})=-\frac{1}{\varepsilon}\sum_{\substack{{\rm even}\\ (i,j,k)}}\xi_j e^{a}_{ii}\otimes e^\ell_{kk}
+\sum_{i\neq j}e^{a}_{ij}\otimes e^\ell_{ji}+u^+\varepsilon^{\beta}\sum_{i=1}^3e^{a}_{ii}\otimes e^\ell_{ii}
-\varepsilon\sum_{\substack{{\rm odd}\\ (i,j,k)}} \frac{1}{\xi_j} e^{a}_{ii}\otimes e^\ell_{kk}.
\end{equation}
If we interpret $\mathbf{\tilde{R}}^{a,\ell}$ as the Boltzmann weights on a vertex with a horizontal line for the auxiliary space and a vertical line 
for the quantum space, the first and the fourth terms of \eqref{eq:RmatnewII} represent crossings of two different states; 
the second term represents reflection while the third one arises when both lines carry identical states.
These Boltzmann weights are given graphically in Figures \ref{Fig:BoltzmannWeight1} and \ref{Fig:BoltzmannWeight2} .

\begin{figure}[t]
\setlength{\unitlength}{.15cm}
\centerline{
\begin{picture}(50,20)
\thicklines
\thicklines
\put(2.5,-5){\line(1,0){55}}
\put(2.5,2){\line(1,0){55}}
\put(2.5,7){\line(1,0){55}}
\put(2.5,12){\line(1,0){55}}
\put(2.5,19){\line(1,0){55}}
\put(2.5,24){\line(1,0){55}}
\put(5,-7.5){\line(0,1){34}}
\put(10,-7.5){\line(0,1){34}}
\put(15,-7.5){\line(0,1){34}}
\put(20,-7.5){\line(0,1){34}}
\put(25,-7.5){\line(0,1){34}}
\put(35,-7.5){\line(0,1){34}}
\put(55,-7.5){\line(0,1){34}}
\put(50,-7.5){\line(0,1){34}}
\put(45,-7.5){\line(0,1){34}}
\put(40,-7.5){\line(0,1){34}}
\put(29,-10){$\cdots$}
\put(4.5,-10){\scriptsize 1}
\put(14.5,-10){\scriptsize 1}
\put(24.5,-10){\scriptsize 1}
\put(34.5,-10){\scriptsize 1}
\put(39.5,-10){\scriptsize 1}
\put(44.5,-10){\scriptsize 1}
\put(49.5,-10){\scriptsize 1}
\put(54.5,-10){\scriptsize 1}
\put(9.5,-10){\scriptsize 1}
\put(19.5,-10){\scriptsize 1}
\put(4,28){\scriptsize $n_1$}
\put(9,28){\scriptsize $n_2$}
\put(14,28){\scriptsize $n_3$}
\put(19,28){\scriptsize $n_4$}
\put(47,28){\scriptsize $n_{L-1}$}
\put(54,28){\scriptsize $n_{L}$}
\put(29,27){$\cdots$}
\put(32.5,27){$\cdots$}
\put(25.5,27){$\cdots$}
\put(36,27){$\cdots$}
\put(39.5,27){$\cdots$}
\put(-1,-5){\scriptsize $b_K$}
\put(-1,-1.7){$\vdots$}
\put(59,-1.7){$\vdots$}
\put(-1,15){$\vdots$}
\put(59,15){$\vdots$}
\put(-12,-1.7){$\vdots$}
\put(-12,15){$\vdots$}
\put(-1,2){\scriptsize $b_2$}
\put(-1,7){\scriptsize $b_1$}
\put(-1,12){\scriptsize $a_{M'}$}
\put(-1,19){\scriptsize $a_2$}
\put(-1,24){\scriptsize $a_1$}
\put(59,24){\scriptsize 1}
\put(59,19){\scriptsize 1}
\put(59,12){\scriptsize 1}
\put(59,7){\scriptsize 1}
\put(59,2){\scriptsize 1}
\put(59,-5){\scriptsize 1}
\put(-15,24){$\mathbf{M}_{1a_1}(u^{\rm I}_1)$}
\put(-15,19){$\mathbf{M}_{1a_2}(u^{\rm I}_2)$}
\put(-15,12){$\mathbf{M}_{a_{M'}}(u^{\rm I}_{M'})$}
\put(-15,7){$\mathbf{M}_{1b_1}(u^{\rm II}_1)$}
\put(-15,2){$\mathbf{M}_{1b_2}(u^{\rm II}_2)$}
\put(-15,-5){$\mathbf{M}_{1b_{K}}(u^{\rm II}_{K})$}
\end{picture}
}
\vskip 1.5cm
\caption{Graph for a state 
$\mathbf{M}_{1a_1}(u^{\rm I}_1)\cdots\mathbf{M}_{1a_{M'}}(u^{\rm I}_{M'})\mathbf{M}_{1b_1}(u^{\rm II}_1)
\cdots\mathbf{M}_{1b_{K}}(u^{\rm II}_{K})\vert 0\rangle$}
\label{Fig:BetheVector}
\end{figure}
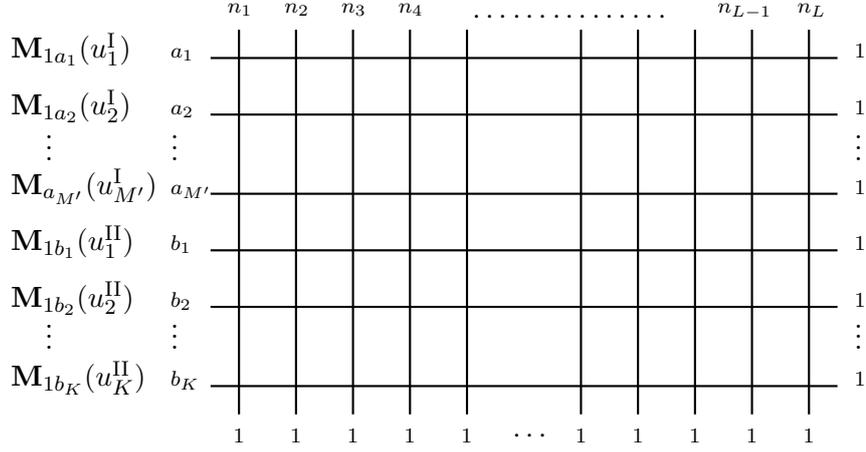

A Bethe state $\mathbf{M}_{1a_1}(u^{\rm I}_1)\cdots\mathbf{M}_{1a_{M'}}(u^{\rm I}_{M'})
\mathbf{M}_{1b_1}(u^{\rm II}_1)\cdots\mathbf{M}_{1b_{K}}(u^{\rm II}_{K})\vert 0\rangle$
may then be represented by a two-dimensional square lattice as shown in
Figure \ref{Fig:BetheVector}. 
Here the indices $a_i$ and $b_j$ are either $2$ or $3$, with the condition that
the total numbers of $2$'s and $3$'s should be $M'$ and $K$, respectively.
The Boltzmann weights imposed on the vertices belonging to the top $M'$ horizontal lines are given
by \eqref{eq:RmatnewI} (Fig.\ \ref{Fig:BoltzmannWeight1}); those belonging to the bottom $K$ lines
by \eqref{eq:RmatnewII} (Fig.\ \ref{Fig:BoltzmannWeight2}).
The configuration $|11\cdots 11\rangle$ at the bottom of the graph defines the reference state
$\vert 0\rangle$; that of the top $\vert n_1 n_2 \cdots n_L\rangle$ defines the Bethe state.
Among $L$ numbers $n_1,\cdots, n_L$, the number of $2$'s should be $M'$, that of $3$'s be $K$, and that of $1$'s be $L-M$, since
the indices $a_i$ and $b_j$ on the left side, which are either $2$ or $3$, should appear 
on the top side since the numbers of $1,2,3$ states are individually  conserved by the Boltzmann weights. (Notice that states on the bottom and the right sides are all $1$'s.)
Therefore, the resulting state should contain exactly the same number of configurations as
\eqref{eq:totalconfig}.

Since the creation operators do not commute, the orderings of the operators in the definition the Bethe state \eqref{eq:ABA1} matter. 
Among possible orderings, we start with a state
$\mathbf{M}_{12}(u^{\rm I}_1)\cdots\mathbf{M}_{12}(u^{\rm I}_{M'})\mathbf{M}_{13}(u^{\rm II}_1)
\cdots\mathbf{M}_{13}(u^{\rm II}_{K})\vert 0\rangle$.
In fact, as we will show later, this is only ordering which can contribute in the $\varepsilon\to 0$ limit.
A generic configuration (and its shifts) in the $(L,M,K)$ sector can be written, for example, as 
\begin{equation}\label{eq:mostconfig}
\mathbf{f}=(\stackrel{\mathbf{f'}}{\overbrace{{\color{blue}2}1{\color{red}3}{\color{blue}2}
\cdots {\color{blue}2}{\color{red}3}{\color{blue}2}1\cdots{\color{red}3}}}11\cdots 11),\qquad
\vert\,\pi^\ell(\mathbf{f})\rangle
=\vert\stackrel{\ell}{\overbrace{1\cdots 1}}\stackrel{\mathbf{f'}}{\overbrace{{\color{blue}2}1{\color{red}3}{\color{blue}2}\cdots {\color{blue}2}{\color{red}3}{\color{blue}2}1\cdots{\color{red}3}}}1\cdots 1\,\rangle.
\end{equation}
We express states $1,\, 2,\, 3$ with black, blue, and red colors, respectively, as well as the lines 
carrying these states in the graphs for better visibility.
We have denoted a sub-configuration $\mathbf{f'}$ in $\mathbf{f}$ 
which is certain sequence of 
all $2$-states, all $3$'s and some $1$ states whose number is $\mathsf{b}$.
The length of $\mathbf{f'}$ is $M+\mathsf{b}$.

\begin{figure}[t]
\setlength{\unitlength}{.15cm}
\centerline{
\begin{picture}(50,30)(10,0)
\thicklines
\put(21,-5){\line(0,1){30}}
\put(51,-5){\line(0,1){30}}
\put(21,25){\line(1,0){30}}
\put(21,-5){\line(1,0){30}}
\color{red}
\put(0,-4){\line(1,0){21}}
\put(0,0){\line(1,0){21}}
\put(0,3){\line(1,0){21}}
\put(0,6){\line(1,0){21}}
\put(39,25){\line(0,1){2.5}}
\put(42,25){\line(0,1){2.5}}
\put(45,25){\line(0,1){2.5}}
\put(50,25){\line(0,1){2.5}}
\color{blue}
\put(0,23){\line(1,0){21}}
\put(0,20){\line(1,0){21}}
\put(0,17){\line(1,0){21}}
\put(0,14){\line(1,0){21}}
\put(0,9){\line(1,0){21}}
\put(22,25){\line(0,1){2.5}}
\put(25,25){\line(0,1){2.5}}
\put(28,25){\line(0,1){2.5}}
\put(31,25){\line(0,1){2.5}}
\put(36,25){\line(0,1){2.5}}
\color{black}
\thinlines
\put(3,8){\dashbox{0.8}(16,16)[s]{}}
\put(3,-5){\dashbox{0.8}(16,12)[s]{}}
\put(53,8){\dashbox{0.8}(13,16)[s]{}}
\put(53,-5){\dashbox{0.8}(13,12)[s]{}}
\put(51,-4){\line(1,0){17}}
\put(51,0){\line(1,0){17}}
\put(51,3){\line(1,0){17}}
\put(51,6){\line(1,0){17}}
\put(51,23){\line(1,0){17}}
\put(51,20){\line(1,0){17}}
\put(51,17){\line(1,0){17}}
\put(51,14){\line(1,0){17}}
\put(51,9){\line(1,0){17}}
\put(22,-7.5){\line(0,1){2.5}}
\put(25,-7.5){\line(0,1){2.5}}
\put(28,-7.5){\line(0,1){2.5}}
\put(31,-7.5){\line(0,1){2.5}}
\put(36,-7.5){\line(0,1){2.5}}
\put(39,-7.5){\line(0,1){2.5}}
\put(42,-7.5){\line(0,1){2.5}}
\put(45,-7.5){\line(0,1){2.5}}
\put(50,-7.5){\line(0,1){2.5}}
\put(4,-7.5){\line(0,1){35}}
\put(7,-7.5){\line(0,1){35}}
\put(10,-7.5){\line(0,1){35}}
\put(15,-7.5){\line(0,1){35}}
\put(18,-7.5){\line(0,1){35}}
\put(54,-7.5){\line(0,1){35}}
\put(57,-7.5){\line(0,1){35}}
\put(62,-7.5){\line(0,1){35}}
\put(65,-7.5){\line(0,1){35}}
\put(58,-10){$\cdots$}
\put(46,-10){$\cdots$}
\put(32,-10){$\cdots$}
\put(11,-10){$\cdots$}
\put(3.5,-10){\scriptsize 1}
\put(6.5,-10){\scriptsize 1}
\put(9.5,-10){\scriptsize 1}
\put(14.5,-10){\scriptsize 1}
\put(17.5,-10){\scriptsize 1}
\put(3.5,28){\scriptsize 1}
\put(6.5,28){\scriptsize 1}
\put(9.5,28){\scriptsize 1}
\put(14.5,28){\scriptsize 1}
\put(17.5,28){\scriptsize 1}
\put(21.5,-10){\scriptsize 1}
\put(24.5,-10){\scriptsize 1}
\put(27.5,-10){\scriptsize 1}
\put(30.5,-10){\scriptsize 1}
\put(35.5,-10){\scriptsize 1}
\put(38.5,-10){\scriptsize 1}
\put(41.5,-10){\scriptsize 1}
\put(44.5,-10){\scriptsize 1}
\put(49.5,-10){\scriptsize 1}
\put(64.5,-10){\scriptsize 1}
\put(53.5,-10){\scriptsize 1}
\put(56.5,-10){\scriptsize 1}
\put(61.5,-10){\scriptsize 1}
\put(64.5,28){\scriptsize 1}
\put(53.5,28){\scriptsize 1}
\put(56.5,28){\scriptsize 1}
\put(61.5,28){\scriptsize 1}
\put(69,-4.6){\scriptsize 1}
\put(69,-0.3){\scriptsize 1}
\put(69,2.5){\scriptsize 1}
\put(69,5.5){\scriptsize 1}
\put(69,8.5){\scriptsize 1}
\put(69,13.5){\scriptsize 1}
\put(69,16.5){\scriptsize 1}
\put(69,19.5){\scriptsize 1}
\put(69,22.5){\scriptsize 1}
\put(46,28){$\cdots$}
\put(32,28){$\cdots$}
\put(58,28){$\cdots$}
\put(11,28){$\cdots$}
\put(-2,-2.8){$\vdots$}
\put(69,-2.8){$\vdots$}
\put(-2,10.7){$\vdots$}
\put(69,10.7){$\vdots$}
\put(-2,-5){\scriptsize ${\color{red}3}$}
\put(-2,-0.3){\scriptsize ${\color{red}3}$}
\put(-2,2.5){\scriptsize ${\color{red}3}$}
\put(-2,5.5){\scriptsize ${\color{red}3}$}
\put(-2,8.5){\scriptsize ${\color{blue}2}$}
\put(-2,13.5){\scriptsize ${\color{blue}2}$}
\put(-2,16.5){\scriptsize ${\color{blue}2}$}
\put(-2,19.5){\scriptsize ${\color{blue}2}$}
\put(-2,22.5){\scriptsize ${\color{blue}2}$}
\put(21.5,28){\scriptsize ${\color{blue}2}$}
\put(24.5,28){\scriptsize $1$}
\put(27.5,28){\scriptsize ${\color{red}3}$}
\put(30.5,28){\scriptsize ${\color{blue}2}$}
\put(35.5,28){\scriptsize ${\color{blue}2}$}
\put(38.5,28){\scriptsize ${\color{red}3}$}
\put(41.5,28){\scriptsize ${\color{blue}2}$}
\put(44.5,28){\scriptsize $1$}
\put(49.5,28){\scriptsize ${\color{red}3}$}
\put(3.7,30){$\stackrel{\ell}{\overbrace{\qquad\qquad\ \ \quad}}$}
\put(53.5,30){$\stackrel{L-M-\ell-\mathsf{b}}{\overbrace{\qquad\qquad\ \ }}$}
\put(22,30){$\stackrel{M+\mathsf{b}\ (\mathbf{f'})}{\overbrace{\hspace{4.2cm}}}$}
\end{picture}
}
\vskip 1.5cm
\caption{Graph for
$\mathbf{M}_{12}(u^{\rm I}_1)\cdots\mathbf{M}_{12}(u^{\rm I}_{M'})\mathbf{M}_{13}(u^{\rm II}_1)
\cdots\mathbf{M}_{13}(u^{\rm II}_{K})\vert 0\rangle$}
\label{Fig:configurations}
\end{figure}
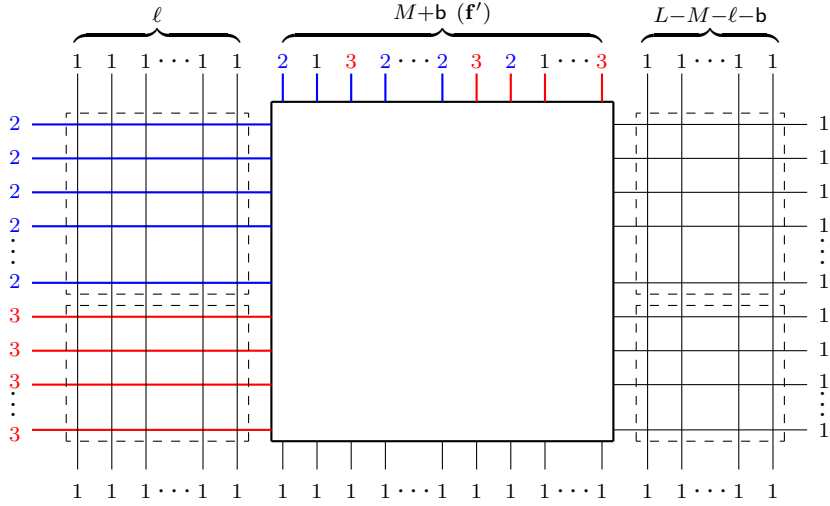

%
The state $\vert\,\pi^\ell(\mathbf{f})\rangle$ is represented graphically in 
Fig.\ \ref{Fig:configurations}.
The blue and red horizontal lines entering from the left side can move only in the right or the upward directions, as one can see from the Boltzmann weights listed above.
Therefore, all blue and red lines should be contained in the large box at the center.
We define $\mathsf{S}(\mathbf{f'})$ for sum of all the weights for many different paths of 
blue, red, and black lines in this box.
Although this factor turns out to be quite complicated, see below, it depends only on $\mathbf{f'}$ and
is obviously independent of $\ell$.

The vertices in other parts of the lattice are ``frozen'' in the sense that the states (colors) 
are all fixed.
While the right part of the box contains only black lines,
the left part contains horizontal color lines, such that only vertical black lines are allowed.
Since states are arranged uniquely, it is straightforward to compute the products of the vertex weights.
Using the Boltzmann weights in Fig.\ \ref{Fig:BoltzmannWeight1} and Fig.\ \ref{Fig:BoltzmannWeight2}, 
we find that the products of the vertex weights in the upper-right, upper-left, lower-right, and lower-left 
dash boxes in Fig.\ \ref{Fig:configurations} are, respectively, given by
\begin{equation}
1,\quad {\displaystyle\prod_{j=1}^{M-K}}\left(\frac{\xi_3 u^-_j}{\varepsilon^{1-\alpha}}\right)^\ell,\quad
\prod_{l=1}^{K}\left(\varepsilon^\beta u^+_l\right)^{L-M-\ell-\mathsf{b}},\quad 
\left(-\frac{\varepsilon}{\xi_2}\right)^{\ell K}.
\end{equation}
Combining, and factoring out constant terms, we obtain\footnote{
This result is not valid if the box extends over the left or right boundaries.
We have computed these boundary cases separately and found that
the result \eqref{eq:BVcomputation1} still holds for the most leading configuration $\mathbf{\bar f}$.}
\begin{equation}\label{eq:BVcomputation1}
\vert\psi(\mathbf{f})\rangle=\left(\varepsilon^{\beta K}\prod_{l=1}^Ku^+_l\right)^{L-M-\mathsf{b}}
\sum_{\ell=1}^{L}\varepsilon^{\ell[(1-\beta)K-(1-\alpha)(M-K)]}
\left[\frac{\xi_3^{M-K}}{(-\xi_2)^K}\frac{{\displaystyle \prod_{j=1}^{M-K}u^-_j}}
{\displaystyle {\prod_{l=1}^{K}u^+_l}}\right]^\ell\vert\,\pi^\ell(\mathbf{f})\,\rangle.
\end{equation}

%
The power of $\varepsilon$ inside the sum, $\ell[(1-\beta)-(1-\alpha)(M-K)]$, vanishes from \eqref{eq:alpha} and \eqref{eq:beta} so that all terms in the sum are of the same order in $\varepsilon$ as required. 
The factor inside the square bracket in \eqref{eq:BVcomputation1} has been identified already as a root of unity $\omega_L^{-k}$ 
in \eqref{eq:transferLMK} and \eqref{eq:explicitEV}.
This proves that the vertices outside the box give the eigenstates of the shift operator
$\vert\mathbf{f}_k\rangle$.
It is important to notice that only this special combination of $\xi_i$'s and the Bethe roots can yield 
this root of unity, and that the other orderings such as 
$\mathbf{M}_{13}(u^{\rm I}_1)\mathbf{M}_{12}(u^{\rm I}_{2})\mathbf{M}_{13}(u^{\rm II}_1)
\mathbf{M}_{12}(u^{\rm II}_{2})\vert 0\rangle$ cannot yield the required
states.\footnote{This is also observed by direct computations and Mathematica code for some simple cases.
}
In summary, the Bethe state in the strong twist limit can be written, up to an $\mathbf{f}$-independent constant, as
\begin{equation}\label{eq:shiftevbasisexpand2}
\vert\psi_{\Lambda,k}\rangle=\sum_{\mathbf{f}}
\left[\,\varepsilon^{-\mathsf{b}\beta K}\mathsf{S}(\mathbf{f'})\,\right]\,
\vert\,\mathbf{f}_k\,\rangle.
\end{equation}
This determines the coefficients $a(\mathbf{f})$ in \eqref{eq:shiftevbasis}.

Now we need to compute $\mathsf{S}(\mathbf{f'})$ from the vertices inside the box. 
Since these are not frozen, there are too many possibilities to compute all their weights in general.
Therefore, we will content ourselves with finding the most dominant configuration for 
a rather simple yet non-trivial case:  $M=3,\, K=1$ with arbitrary $L$.
Based on this result, we will formulate a general conjecture for generic $(L,M,K)$ sectors.

\begin{figure}[t]
\setlength{\unitlength}{.15cm}
\centerline{
\begin{picture}(100,10)(-3,-2)
\thicklines
\put(0,0){\line(0,1){9}}
\put(9,0){\line(0,1){9}}
\put(0,9){\line(1,0){9}}
\put(0,0){\line(1,0){9}}
\color{blue}
\put(-1.5,7.5){\line(1,0){3}}
\put(-1.5,4.5){\line(1,0){3}}
\put(1.5,7.5){\line(1,0){3}}
\put(1.5,4.5){\line(0,1){3}}
\put(4.5,7.5){\line(0,1){3}}
\put(1.5,7.5){\line(0,1){3}}
\color{red}
\put(-1.5,1.5){\line(1,0){3}}
\put(4.5,7.5){\line(1,0){3}}
\put(1.5,4.5){\line(1,0){3}}
\put(1.5,1.5){\line(0,1){3}}
\put(4.5,4.5){\line(0,1){3}}
\put(7.5,7.5){\line(0,1){3}}
\color{black}
\thinlines
\put(1.5,-1.5){\line(0,1){3}}
\put(4.5,-1.5){\line(0,1){6}}
\put(7.5,-1.5){\line(0,1){9}}
\put(7.5,7.5){\line(1,0){3}}
\put(4.5,4.5){\line(1,0){6}}
\put(1.5,1.5){\line(1,0){9}}
\thicklines
\put(20,0){\line(0,1){9}}
\put(29,0){\line(0,1){9}}
\put(20,9){\line(1,0){9}}
\put(20,0){\line(1,0){9}}
\color{blue}
\put(18.5,4.5){\line(1,0){3}}
\put(18.5,7.5){\line(1,0){9}}
\put(21.5,4.5){\line(0,1){3}}
\put(21.5,7.5){\line(0,1){3}}
\put(27.5,7.5){\line(0,1){3}}
\color{red}
\put(18.5,1.5){\line(1,0){3}}
\put(21.5,1.5){\line(0,1){3}}
\put(21.5,4.5){\line(1,0){3}}
\put(24.5,4.5){\line(0,1){3}}
\put(24.5,7.5){\line(0,1){3}}
\color{black}
\thinlines
\put(21.5,-1.5){\line(0,1){3}}
\put(24.5,-1.5){\line(0,1){6}}
\put(27.5,-1.5){\line(0,1){9}}
\put(27.5,7.5){\line(1,0){3}}
\put(24.5,4.5){\line(1,0){6}}
\put(21.5,1.5){\line(1,0){9}}
\thicklines
\multiput(56,-2)(0,1){13}{\line(0,1){0.5}}
\thicklines
\put(40,0){\line(0,1){9}}
\put(49,0){\line(0,1){9}}
\put(40,9){\line(1,0){9}}
\put(40,0){\line(1,0){9}}
\color{blue}
\put(38.5,4.5){\line(1,0){6}}
\put(44.5,4.5){\line(0,1){6}}
\put(38.5,7.5){\line(1,0){9}}
\put(47.5,7.5){\line(0,1){3}}
\color{red}
\put(38.5,1.5){\line(1,0){3}}
\put(41.5,1.5){\line(0,1){9}}
\color{black}
\thinlines
\put(41.5,-1.5){\line(0,1){3}}
\put(44.5,-1.5){\line(0,1){6}}
\put(47.5,-1.5){\line(0,1){9}}
\put(47.5,7.5){\line(1,0){3}}
\put(44.5,4.5){\line(1,0){6}}
\put(41.5,1.5){\line(1,0){9}}
\thicklines
\put(60,0){\line(0,1){9}}
\put(72,0){\line(0,1){9}}
\put(60,9){\line(1,0){12}}
\put(60,0){\line(1,0){12}}
\color{blue}
\put(58.5,4.5){\line(1,0){3}}
\put(58.5,7.5){\line(1,0){6}}
\put(61.5,4.5){\line(0,1){3}}
\put(61.5,7.5){\line(0,1){3}}
\put(64.5,7.5){\line(0,1){3}}
\color{red}
\put(58.5,1.5){\line(1,0){3}}
\put(61.5,1.5){\line(1,0){3}}
\put(64.5,4.5){\line(1,0){3}}
\put(67.5,7.5){\line(1,0){3}}
\put(64.5,1.5){\line(0,1){3}}
\put(67.5,4.5){\line(0,1){3}}
\put(70.5,7.5){\line(0,1){3}}
\color{black}
\thinlines
\put(61.5,-1.5){\line(0,1){6}}
\put(64.5,-1.5){\line(0,1){3}}
\put(64.5,4.5){\line(0,1){3}}
\put(61.5,4.5){\line(1,0){3}}
\put(64.5,7.5){\line(1,0){3}}
\put(67.5,-1.5){\line(0,1){6}}
\put(67.5,7.5){\line(0,1){3}}
\put(70.5,-1.5){\line(0,1){9}}
\put(70.5,7.5){\line(1,0){3}}
\put(67.5,4.5){\line(1,0){6}}
\put(64.5,1.5){\line(1,0){9}}
\thicklines
\put(80,0){\line(0,1){9}}
\put(92,0){\line(0,1){9}}
\put(80,9){\line(1,0){12}}
\put(80,0){\line(1,0){12}}
\color{blue}
\put(78.5,4.5){\line(1,0){6}}
\put(81.5,7.5){\line(0,1){3}}
\put(78.5,7.5){\line(1,0){3}}
\put(84.5,4.5){\line(0,1){3}}
\put(84.5,7.5){\line(1,0){3}}
\put(87.5,7.5){\line(0,1){3}}
\color{red}
\put(78.5,1.5){\line(1,0){3}}
\put(81.5,1.5){\line(1,0){3}}
\put(84.5,4.5){\line(1,0){3}}
\put(87.5,7.5){\line(1,0){3}}
\put(84.5,1.5){\line(0,1){3}}
\put(87.5,4.5){\line(0,1){3}}
\put(90.5,7.5){\line(0,1){3}}
\color{black}
\thinlines
\put(84.5,7.5){\line(0,1){3}}
\put(81.5,-1.5){\line(0,1){9}}
\put(84.5,-1.5){\line(0,1){3}}
\put(87.5,-1.5){\line(0,1){6}}
\put(90.5,-1.5){\line(0,1){9}}
\put(90.5,7.5){\line(1,0){3}}
\put(87.5,4.5){\line(1,0){6}}
\put(84.5,1.5){\line(1,0){9}}
\put(81.5,7.5){\line(1,0){3}}
\put(0,-4){{\scriptsize $({\color{blue}2}{\color{blue}2}{\color{red}3}):\varepsilon^{2\beta}$}}
\put(19,-4){{\scriptsize$({\color{blue}2}{\color{red}3}{\color{blue}2}):\varepsilon^{1+\alpha+2\beta}$}}
\put(39,-4){{\scriptsize$({\color{red}3}{\color{blue}2}{\color{blue}2}):\varepsilon^{2+2\alpha+2\beta}$}}
\put(61,-4){{\scriptsize$({\color{blue}2}{\color{blue}2}1{\color{red}3}):\varepsilon^{1+\beta}$}}
\put(81,-4){{\scriptsize$({\color{blue}2}1{\color{blue}2}{\color{red}3}):\varepsilon^{\alpha+\beta}$}}
\put(3,-7){\scriptsize (a)}
\put(23,-7){\scriptsize (b)}
\put(44,-7){\scriptsize (c)}
\put(65,-7){\scriptsize (d)}
\put(85,-7){\scriptsize (e)}
\end{picture}
}
\vskip 0.7cm
\caption{The vertices inside the box for $M=3,K=1$ for the leading configurations. For (d) and (e), 
$\varepsilon^{-\beta}$ in \eqref{eq:shiftevbasisexpand2} is multiplied. (a) is the most dominant.}
\label{Fig:squareconfig}
\end{figure}
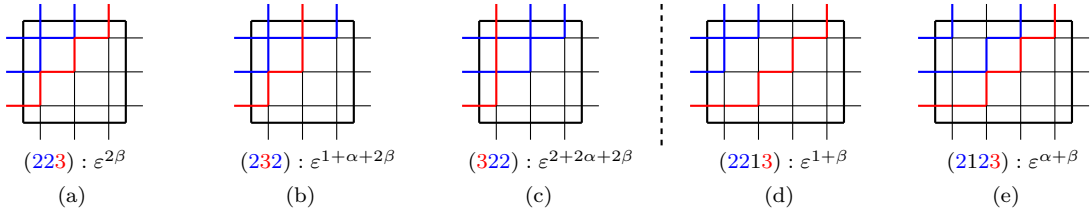

For $\mathsf{b}=0$, the allowed sub-configurations for $M=3,\, K=1$ are
$({\color{blue}2}{\color{blue}2}{\color{red}3}),({\color{blue}2}{\color{red}3}{\color{blue}2}),({\color{red}3}{\color{blue}2}{\color{blue}2})$. 
The vertices that contribute most to these are illustrated in Fig.\ \ref{Fig:squareconfig} (a), (b), and (c), along with the powers of $\varepsilon$ computed from the Boltzmann weights.
From this, we can find that the most dominant $\mathbf{f'}$ is $({\color{blue}2}{\color{blue}2}{\color{red}3})$.

For $\mathsf{b}=1$, we need to consider sub-configurations with $1$ state inserted into 
$({\color{blue}2}{\color{blue}2}{\color{red}3})$; $({\color{blue}2}1{\color{blue}2}{\color{red}3})$ and 
$({\color{blue}2}{\color{blue}2}1{\color{red}3})$.
The vertices for these are illustrated in Fig.\ \ref{Fig:squareconfig} (d) and (e) along with the powers of $\varepsilon$ computed from the Boltzmann weights, where we include an $\varepsilon^{-\mathsf{b}\beta}$  
factor in \eqref{eq:shiftevbasisexpand2}.
Again, we can find that these are sub-leading compared with $({\color{blue}2}{\color{blue}2}{\color{red}3})$
since $1>\alpha>\beta$.
If $\mathsf{b}\ge 2$, more vertices inside the box with two different color lines crossing appear and  
add more positive powers of $\varepsilon$. 
Therefore, these are more suppressed.

Although we have no rigorous proof for the generic cases, 
we conjecture  from this analysis in conjunction with some further specific tests that the most dominant state for general $(L,M,K)$ and cyclicity sector $k$ is given by 
$| \mathbf{\bar f}_k \rangle$ in \eqref{eq:cyclicbasis} with
\begin{equation}\label{eq:locked}
\mathbf{\bar f}=(\stackrel{M-K}{\overbrace{
{\color{blue}2}{\color{blue}2}\cdots{\color{blue}2}}}\,\stackrel{K}
{\overbrace{{\color{red}3}\cdots{\color{red}3}}}\,\stackrel{L-M}{\overbrace{11\cdots 11}} ).
\end{equation}
This implies that the Bethe eigenstates in \eqref{eq:shiftevbasisexpand2} all collapse in the strong twisting limit to
$\vert\,\mathbf{\bar f}_k\,\rangle$, up to a constant of proportionality. We call \eqref{eq:locked} and in fact all associated cyclicity eigenstates $| \mathbf{\bar f}_k \rangle$ {\it locked} states.

This is a disappointing result, considering the model's rich structure of Jordan blocks, see next
chapter \ref{sec:results}.
For each Jordan block, there is exactly one true eigenstate.
However, the limiting process of the standard quantum inverse scattering method yields only one of them per cyclicity sector: A locked state.
This shows that the understanding of the Jordan block structure from integrability needs a new approach; the traditional methodology of the ABA does not suffice.
\section{\mathversion{bold}Jordan Normal Forms for (Hyper)eclectic Spin Chains}
\label{sec:results}
As we argued in the last chapter \ref{sec:QISM}, the ABA does not seem to yield any useful information on the eclectic spin chain models: In the strong twisting limit, all Bethe states collapse to a family of locked states with fixed shift operator eigenvalue, despite the fact that the models stay integrable in the limit. This is closely related to the fact that the Hamiltonian, and actually the entire transfer matrix at non-zero spectral parameter, fail to be diagonalizable in this limit. As already explained in section \ref{sec:failure}, the standard basis change to a diagonal form is to be replaced by a basis change to JNF, see \eqref{eq:energyJNF}, \eqref{eq:transferJNF}. However, it is currently unclear how to find this basis change from integrability. In this chapter, we will demonstrate, by way of examples, that the JNF, while highly intricate, shows very promising regularity properties and seemingly iterative patterns. We will only show results for the Hamiltonian. They were mostly obtained by using the computer algebra programs {\it Wolfram Mathematica} and {\it Matlab}, even though, in a few of the simplest cases, manual calculations based on basic combinatorial considerations suffice. 
Accordingly, the examples of this chapter should be considered experimental and explorative, to be systematically explained in future work.
\subsection{Eclectic Spin Chain}
\label{sec:resultseclectic}

Let us recall the Hamiltonian \eqref{eq:stronglytwistedXXXfinal},\eqref{eq:strongtwistedpermfinal} of the general eclectic model, acting on 
$\underline{\mathbf{3}}^{\otimes L}$.
In an alternative but standard notation, we may also write it as
\begin{equation}\label{eq:Heclectic}
\mathbf{\hat{H}}_{(\xi_1,\xi_2,\xi_3)}=\sum_{\ell=1}^L\left({\xi_3}\, 
\mathbf{e}_{12}^{\ell}\mathbf{e}_{21}^{\ell+1}+
{\xi_1}\, \mathbf{e}_{23}^{\ell}\mathbf{e}_{32}^{\ell+1}+
{\xi_2}\, \mathbf{e}_{31}^{\ell}\mathbf{e}_{13}^{\ell+1}\right),
\end{equation}
where $\mathbf{e}_{12}^{\ell}$ acts only on site $\ell$, with turning a state $2$ into a state $1$ being its only non-zero action, etc. It is clear that this Hamiltonian does not change the length $L$ of the spin chain, nor the numbers $L\!-\!M$ of states $1$, $M\!-\!K$ of states $2$, $K$ of states $3$, respectively. Thus its action is closed on all the states spanned by
\begin{equation}\label{eq:lockedstate}
\vert\stackrel{L-M}{\overbrace{1\cdots 1}}\ \stackrel{M-K}{\overbrace{{\color{blue}2}\cdots {\color{blue}2}}}\
\stackrel{K}{\overbrace{{\color{red}3}\cdots{\color{red}3}}}\rangle
\end{equation}
and all of its possible permutations. We call the associated state space the $(L,M,K)$ sector.
The dimension of this vector space is obviously
\begin{equation}\label{eq:eq:totalconfig2}
\frac{L!}{(L-M)!\,(M-K)!\,K!}\,,
\end{equation}
which, therefore, gives the size of the Hamiltonian matrix in this sector, see also \eqref{eq:totalconfig}.

It is furthermore clear that \eqref{eq:Heclectic} commutes with the shift operator. By a suitable similarity transformation within each $(L,M,K)$ sector, along the lines of \eqref{eq:cyclicbasis}, we may form a new basis spanned by eigenstates of the shift operator with fixed eigenvalue $\omega_L^k$, $k=0,\ldots,L\!-\!1$, see \eqref{eq:rootfunity},\eqref{eq:shiftev}. Thereby the Hamiltonian in \eqref{eq:Heclectic} becomes block-diagonal, with the $L$ blocks\footnote{
Note that the size of a block is roughly but not exactly equal to $\frac{1}{L}\,\frac{L!}{(L-M)!\,(M-K)!\,K!}$. The exact size is determined by Polya counting, see appendix \ref{app:polya}. Related is the fact that the eigenstates to-be \eqref{eq:cyclicbasis} are sometimes zero for some values of $k$.
}
labelled by $k$. Indexing such a block by the label $(L,M,K,k)$, we find from \eqref{eq:energyJNF}, \eqref{eq:energygeneval} that it may be similarity-transformed to the JNF
\begin{equation}\label{eq:energyJNF3}
S \cdot \mathbf{\hat{H}}_{(\xi_1,\xi_2,\xi_3)}^{(L,M,K,k)} \cdot S^{-1}=\left(
\begin{array}{ccc}
 J_{l_1}\left(0\right) &  &      0 \\
    & \ddots &      \\
 0   &    &  J_{l_b}\left(0\right)   
  \end{array}
\right),
\end{equation}
where $J_l(0)$ are Jordan matrices \eqref{eq:JordanBlock} of size $l$ with generalized eigenvalue $0$, $b$ is the total number of Jordan matrices, and their sizes add up to the size $l_1+\ldots + l_b$ of the matrix $\mathbf{\hat{H}}_{(\xi_1,\xi_2,\xi_3)}^{(L,M,K,k)}$. Let us simplify the notation for the rest of this section by defining $J_l:=J_l(0)$, as appropriate for the JNF of the model's Hamiltonian.

In general, the specifics of the Jordan decomposition also depend on the scaled twist parameters $\xi_1,\xi_2,\xi_3$. However, we found that there always exists a {\it generic} JNF for {\it generic} parameters. It holds for ``most'' triplets of parameters. However, by finetuning the parameters different decompositions might appear. Let us give the simplest non-trivial example of this phenomenon:
For $L=3$, $M=2$, $K=1$ one has six states. For generic $\xi_i$'s one has
\begin{equation}\label{eq:H321generic}
S \cdot \mathbf{\hat{H}}_{(\xi_1,\xi_2,\xi_3)}^{(3,2,1)} \cdot S^{-1}=\left(
\begin{array}{ccc}
 J_2 &  &      0 \\
    & J_2 &      \\
 0   &    &  J_2  
  \end{array}
\right)
\quad {\rm with} \quad J_2=\left(
\begin{array}{cc}
0 & 1 \\
0 & 0 
\end{array}\right),
\end{equation}
where the three $2 \times 2$ Jordan blocks correspond to the three cyclicity sectors with $k=0,1,2$. On the other hand, by way of example, if the three twist parameters are all equal, i.e.\ $\xi:=\xi_1=\xi_2=\xi_3$, only the $k=0$ sector stays non-diagonalizable, while the two non-cyclic sectors $k=1,2$ become diagonalizabe. One therefore has instead of \eqref{eq:H321generic}
\begin{equation}\label{eq:H321special}
S \cdot \mathbf{\hat{H}}_{(\xi,\xi,\xi)}^{(3,2,1)} \cdot S^{-1}=
\left(\begin{array}{ccccc}
 J_2 &  &   &   & 0 \\
    & J_1 &   &    &  \\
    &   & J_1 &   &   \\
    &  &   & J_1 &   \\
   0 &   &   &   & J_1
\end{array}\right)
=
\left(\begin{array}{cccccc}
 0 & 1 & 0 & 0 & 0 & 0\\
  0 & 0 & 0 & 0 & 0 & 0\\
   0 & 0 & 0 & 0 & 0 & 0\\
    0 & 0 & 0 & 0 & 0 & 0 \\
     0 & 0 & 0 & 0 & 0 & 0 \\
      0 & 0 & 0 & 0 & 0 & 0 \\
  \end{array}\right).
\end{equation}

\begin{table}[b]
\centerline{
\begin{tabular}{|c|c|l|}
\hline
L& Size of $\mathbf{\hat{H}}_{(\xi_1,\xi_2,\xi_3)}^{(L,2,1,k)}$ &  JNF\\
\hline
5&6  &  1 5 \\
7&15  & 1 5 9 \\
9&28  & 1 5 9 13 \\
\hline
6&10 & 3 7 \\
8&21  &  3 7 11\\
10&36  &  3 7 11 15\\
\hline
\end{tabular}
}
\caption{JNF for $L$, $M=3,K=1$ (numerical analysis).}
\label{table:M2K1}
\end{table}

Before proceeding, let us introduce the concept of a multiset, which provides a useful notation for our JNF. Multisets, unlike sets, allow for multiple occurrences for each of their elements. The union $\cup$ of two multisets means joining them, adding the multiplicities of identical elements. The relative complement $\setminus$ of two multisets is defined in an analogous fashion. In multiset notation, we would describe the Jordan decomposition (all three cyclicity sectors) of \eqref{eq:H321generic} by the multiset $\{2,2,2\}$, while for \eqref{eq:H321special} we would write $\{2,1,1,1,1\}$. An even more concise notation, which we will use extensively below in Tables \ref{table:M4K1}, \ref{table:M5K1} would be $2^3$ for the first 
example, and $1^5 \,2$ for the second example.

Let us next discuss the generalization of \eqref{eq:H321generic} to general $L \geq 3$. It is possible to analytically prove (and easily checked numerically for various values of $L$) that in this case the generic JNF is
\begin{equation}\label{eq:HL21}
S \cdot \mathbf{\hat{H}}_{(\xi_1,\xi_2,\xi_3)}^{(L,2,1)} \cdot S^{-1}=\left(
\begin{array}{ccc}
 J_{L-1} &  &      0 \\
    & \ddots &      \\
 0   &    &  J_{L-1}  
  \end{array}
\right)=(L-1)^L\,
\end{equation}
where the $L$ Jordan blocks of size $L\!-\!1$ correspond to the $L$ cyclicity sectors ($k=0,\ldots L\!-\!1$). 

This simple pattern quickly gets significantly more involved as one increases $M$ and $K$. Let us show some of the emerging structure by fixing $K=1$ (if $K>1$, the complexity of the decompositions further increases). This has the advantage that Polya counting trivializes, as one may consider the position of the single state $3$ as a marker on the spin chain: There is an equal number of states in all $L$ sectors labelled by $(L,M,1,k)$, namely
\begin{equation}\label{eq:eq:totalconfig3}
\frac{(L-1)!}{(L-M)!\,(M-1)!}\,.
\end{equation}
We list the decomposition of $\mathbf{\hat{H}}_{(\xi_1,\xi_2,\xi_3)}^{(L,3,1,k)}$, any cyclicity sector $k$, for $L=5,\ldots,10$ in Table \ref{table:M2K1}. Based on this, it is fairly straightforward to formulate a conjecture for the JNF; it reads, in (multi)set notation,
\begin{equation}\label{eq:conjecture1}
\left\{2L-4\,j-1\ \bigg\vert\ j=1,\cdots,\left[\frac{L-1}{2}\right]\right\}.
\end{equation}
Another point of view to see this are recursion relations.
Let $S^{(M)}_L$ (with $K=1$) be the multiset of the Jordan block sizes for a given $L$ and $M$ in the cyclic ($k=0$) sector. Then, the following relation holds
\begin{equation}\label{eq:recurse1a}
S^{(3)}_{L+2}=S^{(3)}_L \cup \{2L-1\},
\end{equation}
with initial conditions
\begin{equation}\label{eq:recurse1b}
S^{(3)}_1=S^{(3)}_2=\{ \}.
\end{equation}

As $M$ increases, the JNF decomposition patterns get much richer. To uncover these, it is necessary to go to sufficiently high values of $L$. This is not straightforwardly done for the eclectic chain. However, we can get the wanted results for lengths up to $L \sim 20$ for the hypereclectic model, see next section \ref{sec:resultshypereclectic}. As we shall explain, they are then also expected to hold for the generic eclectic model.
\subsection{Hypereclectic Spin Chain}
\label{sec:resultshypereclectic}

The Hamiltonian of the hypereclectic model \eqref{eq:hypereclecticH}, \eqref{eq:hypereclecticP} is obtained from the one of the eclectic chain by setting 
$\xi_1=\xi_2=0$, $\xi_3=1$. In the alternative notation of \eqref{eq:Heclectic} it reads
\begin{equation}\label{eq:Hhypereclectic}
\mathfrak{H}=\sum_{\ell=1}^L \mathbf{e}_{12}^{\ell} \mathbf{e}_{21}^{\ell+1}\,.
\end{equation}
From our exploratory studies at low $L,M,K$ it turns out that this hypereclectic model \eqref{eq:Hhypereclectic} seems to have the {\it same} JNF decomposition as the eclectic model \eqref{eq:Heclectic} for {\it generic} parameters $\xi_1,\xi_2,\xi_3$, as long as the filling conditions \eqref{eq:LMKdomain} are satisfied. (If they are not, the JNF decomposition is different.) Let us call this the {\it universality hypothesis} for the eclectic spin chain models. Let us give one highly non-trivial example, picking $L\!=\!7$. We first consider $M\!=\!3$, $K\!=\!1$. The $15 \times 15$ matrix representations of the respective Hamiltonian matrices in the cyclic $k=0$ sector decompose 
(with different similarity transforms $S$, $S'$, of course) as
\begin{equation}\label{eq:H731generic}
S \cdot \mathfrak{H}^{(7,3,1,k=0)} \cdot S^{-1}=
S' \cdot \mathbf{\hat{H}}_{(\xi_1,\xi_2,\xi_3)}^{(7,3,1,k=0)} \cdot S'^{-1}=\left(
\begin{array}{ccc}
 J_9 &  &      0 \\
    & J_5 &      \\
 0   &    &  J_1  
  \end{array}
\right).
\end{equation}
In the multiset notation introduced in the previous section, this JNF decomposition reads
\begin{equation}\label{eq:multiset0}
\{9,5,1\}=1\,5\,9\,.
\end{equation}
For the eclectic model, this is valid for ``most'' values of $\xi_1,\xi_2,\xi_3$, {\it including} the hypereclectic case $\xi_1=\xi_2=0$, $\xi_3=1$. If we permute the three states $1$, ${\color{blue}2}$, ${\color{red}3}$, this decomposition will 
remain true for the generic eclectic model for all six permutations of these states due to symmetry. On the other hand, the hypereclectic model behaves very different under permutations of the states. The reason is, that the dynamics only involves the states $1$ and ${\color{blue}2}$, while ${\color{red}3}$ is just an inert ``spectator'', forming some kind of wall. So it makes a significant difference whether ${\color{red}3}$ is the least numerous state, second least numerous state, or else the most numerous state. In our specific $L=7$ example, for the cyclic sector, we have again
\begin{equation}\label{eq:multiset1}
\{9,5,1\}=1\,5\,9
\quad \textrm{for permutations of}\,\,\,
\vert 1111{\color{blue}2} {\color{blue}2} {\color{red}3} \rangle\,\, 
\textrm{and}\,\,
\vert {\color{blue}2} {\color{blue}2} {\color{blue}2} {\color{blue}2}11 {\color{red}3} \rangle 
,
\end{equation}
with, respectively, $M\!=\!3$, $K\!=\!1$ and $M\!=\!5$, $K\!=\!1$.
On the other hand, we find the JNF
\begin{equation}\label{eq:multiset2}
\{5,4,3,2,1\}=1\,2\,3\,4\,5
\quad \textrm{for permutations of}\,\,\,
\vert 1111{\color{red}3} {\color{red}3} {\color{blue}2} \rangle\,\, 
\textrm{and}\,\,
\vert {\color{blue}2} {\color{blue}2} {\color{blue}2} {\color{blue}2} {\color{red}3} {\color{red}3} 1 \rangle 
,
\end{equation}
corresponding, respectively, to $M\!=\!3$, $K\!=\!2$ and $M\!=\!6$, $K\!=\!2$. Finally, one has the JNF
\begin{equation}\label{eq:multiset3}
\{3,2,2,2,1,1,1,1,1,1\}=1^6\, 2^3\, 3
\quad \textrm{for permutations of}\,\,\,
\vert {\color{red}3} {\color{red}3} {\color{red}3} {\color{red}3} 1 1 {\color{blue}2} \rangle\,\, 
\textrm{and}\,\,
\vert {\color{red}3} {\color{red}3} {\color{red}3} {\color{red}3} {\color{blue}2} {\color{blue}2} 1 \rangle
,
\end{equation}
with, respectively, $M\!=\!5$, $K\!=\!4$ and $M\!=\!6$, $K\!=\!4$. 

In conclusion, the JNF decompositions of the hypereclectic chain are definitely not invariant under a permutation of the states. This means, its ``spectrum'' is much richer than the one of the generic eclectic chain, despite its simpler looking Hamiltonian.

The hypereclectic model is simple enough to allow for the derivation of a few exact results by some straightforward combinatorics. Let us present one example. Consider cyclic eigenstates with $K=1$, i.e.
\begin{equation}\label{eq:cyclicstate}
\vert 1\cdots 1{\color{blue}2}
1\cdots 1{\color{blue}2}\cdots{\color{blue}2}
1\cdots 1{\color{blue}2}
1\cdots 1
{{\color{red}3}}
\rangle \rangle
:=
\sum_{\ell=1}^L 
\vert 1\cdots 1{\color{blue}2}
1\cdots 1{\color{blue}2}\cdots{\color{blue}2}
1\cdots 1{\color{blue}2}
1\cdots 1
\stackrel{\stackrel{[\ell]}{\downarrow}}{{\color{red}3}}1\cdots 1
\rangle.
\end{equation}
Acting with the Hamiltonian $\mathfrak{H}$ on this state, any ${\color{blue}2}$ situated to the left of a $1$ exchanges with it, thereby moving one step to the right.
None of the remaining $1$'s, ${\color{blue}2}$'s and none of the ${\color{red}3}$'s can move. In a hopefully intuitive notation we have
\begin{eqnarray}\label{eq:intuitive}
\vert 1 \cdots 1
\stackrel{\stackrel{}{\rightarrow}}{ {\color{blue}2}}
1\cdots 1
\stackrel{\stackrel{}{\rightarrow}}{ {\color{blue}2}}
\cdots\stackrel{\stackrel{}{\rightarrow}}{ {\color{blue}2}}
1\cdots 1
\stackrel{\stackrel{}{\rightarrow}}{ {\color{blue}2}}
1\cdots 1
{{\color{red}3}}\cdots 
\rangle \rangle
\end{eqnarray}
Clearly, one can immediately write down an eigenstate that trivially annihilated by the action of the Hamiltonian, a ``locked state'', cf.\ \eqref{eq:locked}:
%
\begin{equation}\label{eq:Jtop}
\mathfrak{H}\, \vert\stackrel{L-M}{\overbrace{11\cdots 11}}
\stackrel{M-1}{\overbrace{
{\color{blue}2}{\color{blue}2}\cdots{\color{blue}2}}}
{\color{red}3}\rangle \rangle=0.
\end{equation}
Here it is easy to write down the lowest Jordan descendent:
\begin{equation}\label{eq:Jdescendent}
\vert\stackrel{M-1}{\overbrace{
{\color{blue}2}{\color{blue}2}\cdots{\color{blue}2}}}
\stackrel{L-M}{\overbrace{11\cdots 11}}{\color{red}3}\rangle \rangle.
\end{equation}
By acting $\mathcal{H}$ on this descendent $l$ times, we obtain the locked state:
\begin{equation}\label{eq:Jordanstring}
\mathfrak{H}^l\, \vert\stackrel{M-1}{\overbrace{
{\color{blue}2}{\color{blue}2}\cdots{\color{blue}2}}}
\stackrel{L-M}{\overbrace{11\cdots 11}}{\color{red}3}\rangle \rangle=
\vert\stackrel{L-M}{\overbrace{11\cdots 11}}
\stackrel{M-1}{\overbrace{
{\color{blue}2}{\color{blue}2}\cdots{\color{blue}2}}}
{\color{red}3}\rangle \rangle,
\end{equation}
where $l$ is found to be $l=(M-1)(L-M)$.
Hence, the size of the largest Jordan block should be given by 
\begin{equation}\label{eq:largestblock}
(M-1)(L-M)+1\,.
\end{equation}
Let us end the section by verifying this numerically for $M=4,5$ and $K=1$. 

In order to reach sufficiently high values of $L$, a direct numerical approach that builds up the hypereclectic Hamiltonian matrix at fixed $M$, $K$ for the cyclic sector is quite feasible up to values of about $L \sim 20$. However, built-in algorithms that subsequently reduce the matrix to JNF are already taking too much computational time. Luckily, a result from Linear Algebra comes in handy. Let $H$ be a matrix over the complex numbers whose generalized eigenvalues are all zero. Let
\begin{equation}\label{eq:dimker1}
a_s:=\mathrm{dim}\, \mathrm{ker}\, H^s
\quad \text{with} \quad
s=0, 1, 2, \ldots\,,
\end{equation}
i.e.\ $a_s$ is the dimension of the kernel of $H^s$.
One then has
\begin{eqnarray}\label{eq:dimker2}
&& a_0=0\,, \qquad a_1=\text{Number of Jordan blocks of}\,\, H\,, \\
&& 2\,a_s-a_{s-1}-a_{s+1}= \text{Number of Jordan blocks of size}\,\, s\, \text{of}\,\, H\,.
\end{eqnarray}
Knowing the $a_s$ clearly determines the structure of the Jordan normal form decomposition of $H$. And \eqref{eq:dimker1} allows to efficiently compute the $a_s$ by Gaussian elimination on a computer, wherewith one brings $H^s$ into echelon form. Clearly we may restrict $s$ to be at most the size of $H$ plus one.

The result of this procedure for $M=4, K=1$ is summarized for $L=6, \ldots 21$ in Table~\ref{table:M4K1}. 
One noticeable feature is that some Jordan blocks are repeating several 
times with definite multiplicities denoted as exponents in the table.
One may again conjecture some simple recursion relations, in generalization of \eqref{eq:recurse1a}, \eqref{eq:recurse1b}.
Let $S^{(4)}_L$ be the multiset of the Jordan block sizes for a given $L$ with $M=4$ and $K=1$. Then, the following relation appears to hold in general:
\begin{equation}\label{eq:recurse2a}
S^{(4)}_{L+4}=S^{(4)}_L \cup \{L+2\,j+1\,\vert\, j=0,\cdots,L-2,L\},
\end{equation}
with initial conditions
\begin{equation}\label{eq:recurse2b}
S^{(4)}_2=S^{(4)}_3=\{\},\quad S^{(4)}_4=\{1\},\quad S^{(4)}_5=\{4\}.
\end{equation}
On easily verifies the result \eqref{eq:largestblock} for the size of the largest Jordan block for all entries of Table~\ref{table:M4K1}.

By the same method, one finds the Jordan block decompositions for $M=5$, $K=1$. The emerging structure is even more intricate,
as shown in Table \ref{table:M5K1} for
$L=8, \ldots\,,18$.
If we denote the multisets of block sizes for a given $L$ with $M=5$, $K=1$ as $S^{(5)}_L$, we find 
the following recursion relation:
\begin{equation}\label{eq:recurse3}
S^{(5)}_{L+5} \setminus S^{(5)}_{L+2}=\left( S^{(5)}_{L+3} \setminus S^{(5)}_{L} \right) \cup
\{\,2L+2j+1\,|\,j=0,\cdots,L-2,L\}.
\end{equation}
Once again one may verify the result \eqref{eq:largestblock} for the size of the largest Jordan block for all entries of Table~\ref{table:M5K1}.

{\renewcommand{\arraystretch}{.9}
\begin{table}[t]
\begin{center}
\begin{tabular}{| c |l|}
\hline
{\scriptsize $L$}&{\scriptsize  Sizes of Jordan Blocks}\\
\hline
{\scriptsize 6}&{\scriptsize 3 7} \\
{\scriptsize 10}&{\scriptsize 3 $7^2$ 9 11 \ 13  \ \ 15  \qquad 19}  \\
{\scriptsize 14}&{\scriptsize 3 $7^2$ 9 $11^2$ $13^2$ $15^2$ 17\ \  $19^2$ 21\ \ \  23\ \  25\ \ \  27 \quad \ \ 31 }  \\
{\scriptsize 18}&{\scriptsize 3 $7^2$ 9 $11^2$ $13^2$ $15^2$ $17^2$ $19^2$ $21^2$ $23^2$ $25^2$ $27^2$ 29 $31^2$ 33 35 37 39 43} \\
\hline
{\scriptsize 8}&{\scriptsize 1 5 7 9\qquad\ 13} \\
{\scriptsize 12}&{\scriptsize 1 5 7 $9^2$ 11 $13^2$ 15\  \, 17\ \, 19\ \,  21\qquad \, 25 }\\
{\scriptsize 16}&{\scriptsize 1 5 7 $9^2$ 11 $13^3$ $15^2$ $17^2$ $19^2$ $21^2$ 23\ \, $25^2$ 27\ \,  29\ \,  31\ \,  33\qquad\ 37 }\\
{\scriptsize 20}&{\scriptsize 1 5 7 $9^2$ 11 $13^3$ $15^2$ $17^3$ $19^3$ $21^3$ $23^2$ $25^3$ $27^2$ $29^2$ $31^2$ $33^2$ 35 $37^2$ 39 41 43 45 49} \\
\hline
{\scriptsize 7}&{\scriptsize 4 6\ \ \,  10} \\
{\scriptsize 11}&{\scriptsize 4 6 8 $10^2$ 12\ \,  14\ \,  16\ \,  18\qquad\,  22 } \\
{\scriptsize 15}&{\scriptsize 4 6 8 $10^2$ $12^2$ $14^2$ $16^2$ $18^2$ 20\ \,  $22^2$\,  24\ \, 26\ \  28 \,  30\qquad\ 34 } \\
{\scriptsize 19}&{\scriptsize 4 6 8 $10^2$ $12^2$ $14^2$ $16^3$ $18^3$ $20^2$ $22^3$ $24^2$ $26^2$ $28^2$ $30^2$ 32 $34^2$ 36 38 40 42 46 }\\
\hline
{\scriptsize 9}&{\scriptsize 4 6 8 10\ \,   12\qquad\ \, 16 }\\
{\scriptsize 13}&{\scriptsize 4 6 8 $10^2$ $12^2$ 14\ \,   $16^2$ 18\ \,   20\ \,   22\ \,   24\qquad\ \,   28  }\\
{\scriptsize 17}&{\scriptsize 4 6 8 $10^2$ $12^2$ $14^2$ $16^3$ $18^2$ $20^2$ $22^2$ $24^2$ 26\ \,   $28^2$ 30\ \,   32 \,  34\ \,   36\qquad 40} \\
{\scriptsize 21}&{\scriptsize 4 6 8 $10^2$ $12^2$ $14^2$ $16^3$ $18^3$ $20^3$ $22^3$ $24^3$ $26^2$ $28^3$ $30^2$ $32^2$ $34^2$ $36^2$ 38 $40^2$ 42 44 46 48 52} \\
\hline
\end{tabular}
\end{center}
\vskip -.4cm
\caption{Structures of Jordan blocks for the sector of $M=4,K=1$, $k=0$ (cyclic states). Exponents denote multiplicities.}
\label{table:M4K1}
\end{table}

%
%
{\renewcommand{\arraystretch}{.9}
\begin{table}[t]
\centerline{
\begin{tabular}{| c |l | }
\hline
{\scriptsize $L$}&{\scriptsize  Sizes of Jordan Blocks}\\
\hline
{\scriptsize 8}&{\scriptsize 1\ \,  5\ \,  7\ \,  9\qquad\ \  13}\\
{\scriptsize 9}&{\scriptsize 1\ \,  $5^2$\ \, \ \,  $9^2$ 11\ \,  13\qquad\ \   17} \\
{\scriptsize 10}&{\scriptsize 1\ \,  $5^2$ 7\ \,  $9^2$ 11\ \,  $13^2$ 15\ \,  17\qquad\, 21} \\
{\scriptsize 11}&{\scriptsize $1^2$ $5^2$ 7\ \,  $9^3$ 11\ \,  $13^3$ 15\ \,  $17^2$ 19\ \,  21\qquad \,  25} \\
{\scriptsize 12}&{\scriptsize 1\ \,  $5^3$ 7\ \,  $9^3$ $11^2$ $13^3$ $15^2$ $17^3$ 19\ \,  $21^2$  23\ \,  25\qquad\, 29} \\
{\scriptsize 13}&{\scriptsize $1^2$ $5^3$ 7\ \,  $9^4$ $11^2$ $13^4$ $15^2$ $17^4$ $19^2$ $21^3$ 23\ \,  $25^2$ 27\ \,  29\qquad\, 33} \\
{\scriptsize 14}&{\scriptsize $1^2$ $5^3$ $7^2$ $9^4$ $11^2$ $13^5$ $15^3$ $17^4$ $19^3$ $21^4$ $23^2$ $25^3$ 27\ \,  $29^2$ 31\ \,  33
\qquad\ 37 }\\
{\scriptsize 15}&{\scriptsize $1^2$ $5^4$ 7\ \,  $9^5$ $11^3$ $13^5$ $15^3$ $17^6$ $19^3$ $21^5$ $23^3$ $25^4$ $27^2$ $29^3$ 31\ \, 
 $33^2$ 35\ \,  37\qquad\ \  41}  \\
{\scriptsize 16}&{\scriptsize $1^2$ $5^4$ $7^2$ $9^5$ $11^3$ $13^6$ $15^4$ $17^6$ $19^4$ $21^6$ $23^4$ $25^5$ $27^3$ $29^4$ $31^2$ $33^3$ 
35\ \,  $37^2$ 39\ \,  41\qquad 45} \\
{\scriptsize 17}&{\scriptsize $1^3$ $5^4$ $7^2$ $9^6$ $11^3$ $13^7$ $15^4$ $17^7$ $19^5$ $21^7$ $23^4$ $25^7$ $27^4$ $29^5$ $31^3$ $33^4$ $35^2$ $37^3$ 39\ \, $41^2$ 43 45\qquad 49} \\
{\scriptsize 18}&{\scriptsize $1^2$ $5^5$ $7^2$ $9^6$ $11^4$ $13^7$ $15^5$ $17^8$ $19^5$ $21^8$ $23^6$ $25^7$ $27^5$ $29^7$ $31^4$ $33^5$ $35^3$ $37^4$ $39^2$ $41^3$ 43 $45^2$ 47 49 53}\\
\hline
\end{tabular}
}
\caption{Structures of Jordan blocks for the sector of $M=5,K=1$, $k=0$ (cyclic states). Exponents denote multiplicities.}
\label{table:M5K1}
\end{table}
\section{\mathversion{bold}Conclusions and Open Questions}
\label{sec:conclusions}

We have begun the systematic study of a class of non-diagonalizable, integrable, chiral spin chains which were christened {\it eclectic spin chains} in \cite{Ipsen:2018fmu}. They still contain three free, complex twist parameters $(\xi_1, \xi_2, \xi_3)$, and were 
originally inspired by parts of the one-loop dilatation operator of a strongly twisted, double-scaled deformation of $\mathcal{N}=4$ Super Yang-Mills Theory
\cite{Gurdogan:2015csr,Sieg:2016vap,Caetano:2016ydc,Chicherin:2017cns,Gromov:2017cja,Chicherin:2017frs,Grabner:2017pgm,Kazakov:2018hrh, Gromov:2018hut}. However, here we systematically study these models on their own right, without further exploring their relation to gauge theory. We also introduced a seemingly even simpler version of these models, which we called {\it hypereclectic spin chain}.

Being non-diagonalizable, the goal is to bring Hamiltonian and transfer matrix of these models into JNF. We found ample evidence for a highly intricate yet subtly structured ``spectrum'' of Jordan blocks, in dire need of a systematic understanding. Interestingly, the spectrum of the hypereclectic chain is richer than the one of the eclectic model at generic twist parameters $(\xi_1, \xi_2, \xi_3)$, even though the first model possesses a simpler Hamiltonian. 

A puzzling aspect is that, despite the easily demonstrated integrability of these models, the traditional means of the quantum inverse scattering method appear to fail to describe these models' Jordan decompositions. We demonstrated this in some detail for the ABA method. Particularly vexing is the fact that the BAE remain sensible in the limit, and exhibit a mathematically rich set of solutions that even lead to the correct counting of states. However, the associated Bethe states collapse to a small set of locked states that are, apparently, essentially useless for finding the spectrum of Jordan blocks. 

To summarize, the ultimate goal for the future is then to solve the eclectic models, i.e.\ to find their intricate spectrum of Jordan blocks, by using integrability. Here we would like to draw attention to \cite{Gainutdinov:2016pxy}, where this was understood for a different non-diagonalizable spin chain model. However, the model is quite different; in particular, it only contains blocks of size one or two.

In our case, it is not clear which model will be ``easier'' to treat: The eclectic or hypereclectic one? It is also not obvious whether it is best to concentrate on their scaled R-matrix, or to better proceed from the finitely twisted R-matrix in conjunction with a suitable limiting procedure. Furthermore, is it better to concentrate on the models' Hamiltonians, or else on their commuting transfer matrices? Finally, could it be that these models may be solved by some suitable combinatorial methods, thereby bypassing the power of integrability?
\section*{Acknowledgments}
\label{sec:acknowledgements}
We are very thankful to Luke Corcoran and Leo Zippelius for many helpful discussions, and to Luke for useful comments on the draft.
We would like to express our sincere gratitude to  the {\it Brain Pool Program} of the {\it Korean National Research Foundation} (NRF) under grant {\bf 2-2019-1283-001-1} for generous support of this research. MS thanks Ewha Womans University for hospitality, and CA thanks Humboldt-Universität zu Berlin. 
This work is supported in part by NRF grant (NRF- 2016R1D1A1B02007258) (CA).
\appendix
\section{\mathversion{bold}P\'olya Counting of Cyclic States}
\label{app:polya}

As a consistency check of our rather non-trivial solutions of the scaled BAE for generic $(L,M, K)$ sectors with $L>3(M-K)$, see section \ref{sec:LMK}, we would like to verify that the counting of cyclic states is combinatorially consistent. The correct combinatorics is encoded in the famous P\'olya enumeration theorem.
The latter allows to count the number of inequivalent ``necklaces'' of length $L$ made of ``beads'' of $n$ distinct ``colors''. 
For our case we have $n=3$, with $L\!-\!M$ beads of color $1$, $M\!-\!K$ beads of color ${\color{blue}2}$, and $K$ beads of color ${\color{red}3}$. The number of distinct, cyclically symmetric configurations $d(L,M,K)$ is then found from the  generating function
\begin{equation}\nonumber
Z(x,y,z)=-\sum_{n=1}^{\infty}\frac{\phi(n)}{n}\log\left[1-x^n-y^n-z^n\right]=
\sum_{\substack{L,M,K \\ L\ge M\ge K} } d(L,M,K)\cdot x^{L-M}y^{M-K}z^K,
\end{equation}
defined with the help of Euler's totient function.\footnote{Also known as Euler's phi function. $\phi(n)$ counts the positive integers up to a given integer $n$ that are relatively prime to n.
}
The consistency check then involves testing whether the multiplicities $d(L,M,K)$ match with the degeneracies of the eigenvalues $\omega_L^k=1$ in \eqref{eq:explicitEV}, which were obtained form the Bethe ansatz. This is done by counting the 
number of configurations $\{i_j\},\ \{n_l\}$ in \eqref{eq:explicitEV}.
For the cyclic states, it is the number of cases that give $k=0\ {\rm mod}\ L$. While we did not bother to analyze this analytically, we made an extensive numerical comparison, see Table \ref{table:Polya}, finding a perfect match in all cases considered. We also found it interesting to compare to a ``naive'', only approximately true counting of cyclic states, obtained by dividing the total configuration number \eqref{eq:totalconfig} by $L$.

{\renewcommand{\arraystretch}{.75}
\begin{table}[h]
\begin{center}
\begin{tabular}{| c |c |c|c|c|c| }
\hline
{\scriptsize $L$}&{\scriptsize  $M$}&{\scriptsize $K$}&{\scriptsize naive counting}&{\scriptsize P\'olya counting}&{\scriptsize Bethe ansatz}\\
\hline
{\scriptsize 14}&{\scriptsize 6}&{\scriptsize 2}&{\scriptsize 6435/2}&{\scriptsize 3225}&{\scriptsize 3225}\\
{\scriptsize 16}&{\scriptsize 6}&{\scriptsize 2}&{\scriptsize 15015/2}&{\scriptsize 7518}&{\scriptsize 7518}\\
{\scriptsize 18}&{\scriptsize 6}&{\scriptsize 2}&{\scriptsize 15470}&{\scriptsize 15484}&{\scriptsize 15484}\\
{\scriptsize 20}&{\scriptsize 6}&{\scriptsize 2}&{\scriptsize 29070}&{\scriptsize 29088}&{\scriptsize 29088} \\
{\scriptsize 20}&{\scriptsize 8}&{\scriptsize 2}&{\scriptsize 176358}&{\scriptsize 176400}&{\scriptsize 176400} \\
{\scriptsize 20}&{\scriptsize 10}&{\scriptsize 4}&{\scriptsize 1939938}&{\scriptsize 1940064}&{\scriptsize 1940064} \\
{\scriptsize 21}&{\scriptsize 9}&{\scriptsize 3}&{\scriptsize 1175720}&{\scriptsize 1175730}&{\scriptsize 1175730} \\
{\scriptsize 22}&{\scriptsize 6}&{\scriptsize 2}&{\scriptsize 101745/2}&{\scriptsize 50895}&{\scriptsize  50895}\\
{\scriptsize 22}&{\scriptsize 8}&{\scriptsize 2}&{\scriptsize 406980}&{\scriptsize 407040}&{\scriptsize  407040}\\
{\scriptsize 22}&{\scriptsize 10}&{\scriptsize 4}&{\scriptsize 6172530}&{\scriptsize 6172740}&{\scriptsize 6172740} \\
{\scriptsize 24}&{\scriptsize 6}&{\scriptsize 2}&{\scriptsize 168245/2}&{\scriptsize 84150}&{\scriptsize 84150} \\
{\scriptsize 24}&{\scriptsize 8}&{\scriptsize 2}&{\scriptsize 1716099/2}&{\scriptsize 858132}&{\scriptsize 858132} \\
{\scriptsize 24}&{\scriptsize 9}&{\scriptsize 3}&{\scriptsize 4576264}&{\scriptsize 4576278}&{\scriptsize 4576278} \\
{\scriptsize 24}&{\scriptsize 10}&{\scriptsize 4}&{\scriptsize 17160990}&{\scriptsize 17161320}&{\scriptsize 17161320} \\
\hline
\end{tabular}
\end{center}
\vskip -.4cm
\caption{Agreement of the Bethe ansatz solutions of sec.\ \ref{sec:LMK} with P\'olya counting}
\label{table:Polya}
\end{table}

\bibliography{literature}{}
\bibliographystyle{utphys}
\end{document}